\RequirePackage{fix-cm}

\documentclass[twocolumn,epjc3]{svjour3}  

\smartqed  

\RequirePackage{graphicx}
\usepackage{liealg}
\usepackage{braket}
\usepackage{amsmath}
\usepackage{amssymb}
\usepackage{mathtools}

\RequirePackage[numbers,sort&compress]{natbib}
\RequirePackage[colorlinks,citecolor=blue,urlcolor=blue,linkcolor=blue]{hyperref}

\newcommand{\ketjh}[2]{\left|\begin{array}{c}
#1 \\
#2 \\
\end{array}\right\rangle}

\newcommand{\rketjh}[2]{\left\|\begin{array}{c}
#1 \\
#2 \\
\end{array}\right\rangle}
\newcommand{\rbrajh}[2]{\left\langle\begin{array}{c}
#1 \\
#2 \\
\end{array}\right\|}
\newcommand{\overlapjh}[4]{\left\langle\begin{array}{c}
#1 \\
#2 \\
\end{array}\right|\left.\begin{array}{c}
#3 \\
#4 \\
\end{array}\right\rangle}
\newcommand{\wig}[6]{\left\langle\begin{array}{cc}
#1 & #3 \\
#2 & #4 \\
\end{array}\right|\left.\begin{array}{c}
#5 \\
#6 \\
\end{array}\right\rangle}
\newcommand{\rwig}[6]{\left\langle\begin{array}{cc}
#1 & #3 \\
#2 & #4 \\
\end{array}\right\|\left.\begin{array}{c}
#5 \\
#6 \\
\end{array}\right\rangle}

\usepackage{xcolor}

\journalname{Eur. Phys. J. A}

\begin{document}

\sloppy

\title{A modern Fortran library for $\grpsu{3}$ coupling and recoupling coefficients
}

\author{Jakub Herko\thanksref{e1,addr:nd,addr:triumf}
        \and
        Mark A. Caprio\thanksref{addr:nd}
        \and
        Anna E. McCoy\thanksref{addr:nd,addr:anl,addr:int}
        \and
        Patrick J. Fasano\thanksref{addr:nd,addr:anl,presaddr:ns}
}

\thankstext{e1}{e-mail: jherko@triumf.ca}

\institute{Department of Physics and Astronomy, University of Notre Dame, Notre Dame, Indiana 46556-5670, USA \label{addr:nd}
           \and
           TRIUMF, Vancouver, British Columbia V6T 2A3, Canada \label{addr:triumf}
           \and
           Physics Division, Argonne National Laboratory, Argonne, Illinois 60439-4801, USA \label{addr:anl}
           \and
	   Institute for Nuclear Theory, University of Washington, Seattle, Washington 98195-1550, USA \label{addr:int}
           \and
           \emph{Present Address:} NextSilicon Inc., Minneapolis, Minnesota 55402-1572, USA\label{presaddr:ns}
}

\date{Received: date / Accepted: date}

\maketitle

\begin{abstract}
The group $\grpsu{3}$ has applications in several branches of physics. Many of these applications depend on availability of $\grpsu{3}$ coupling and recoupling coefficients. We have developed a modern Fortran library for calculation of the coupling coefficients, for both the $\grpsu{3}\supset\grpu{1}\times\grpsu{2}$ and $\grpsu{3}\supset\grpso{3}$ group chains, and the recoupling coefficients. The library implements the algorithms of Draayer, Akiyama, and Millener, which are laid out in the paper. Performance of the library has been tested and compared to the Akiyama-Draayer (AD) library implementing the same algorithms as well as to a more recent implementation. Our library works for a larger range of $\grpsu{3}$ quantum numbers and provides more accurate coupling coefficients with large quantum numbers than the AD library.
\end{abstract}

\noindent
{\bf Program Summary and Specifications}\\
\begin{normalsize}
\noindent
\emph{Program title:} \texttt{ndsu3lib}\\
\emph{Licensing provisions:} MIT\\
\emph{Programming language:} Fortran 2003 (with C/C++ headers provided)\\
\emph{Repository and DOI:}\\
\href{https://github.com/nd-nuclear-theory/ndsu3lib.git}{https://github.com/nd-nuclear-theory/ndsu3lib.git}\\
\href{https://doi.org/10.5281/zenodo.16655521}{https://doi.org/10.5281/zenodo.16655521}\\
\emph{Description of problem:} Computation of $\grpsu{3}$ coupling and recoupling coefficients.\\
\emph{Method of solution:} The library implements algorithms of Draayer, Akiyama, and Millener.\\
\emph{Additional comments:} This code depends on external libraries for dense linear algebra (LAPACK), $\grpsu{2}$ coupling and recoupling coefficients (GSL or WIGXJPF), and, optionally, multiprecision floating-point calulations (MPFUN2020). \end{normalsize}

\section{Introduction}
\label{sec:introduction}

Applications of the $\grpsu{3}$ symmetry group arise in, \textit{e.g.}, nuclear physics~\cite{elliottI,elliottII,elliottIII,elliottIV,elliott1999,harvey1968:su3-shell,akiyama2,rowe,rowe89,rowe2,npa-318-1979-1-Hecht,vargas,vargas2,dytrych2007:sp-ncsm-dominance,dytrych2007:sp-ncsm-evidence,dytrych2008:sp-ncsm,dytrych2008:sp-ncsm-deformation,dytrych2013,dytrych2020:emergent-symmetry,ppnp-89-2016-101-Launey,sargsyan,irrepfamilies,mccoy2018:diss}, particle physics~\cite{neeman,pr-125-1962-1067-gellmann,gellman2,gellman,han,lipkin,gross,fritzsch,goity}, and quantum optics~\cite{sanders,tan,klimov,bagaev,alodzhants,alodjants}. In particular, the canonical group chain $\grpsu{3}\supset\grpu{1}\times\grpsu{2}$ appears in problems with flavor degrees of freedom, while the angular momentum group chain $\grpsu{3}\supset\grpso{3}$ plays an important role in nuclear physics.

In such applications, the basis used for calculations is expressed in terms of irreducible representations (irreps) of $\grpsu{3}$, and operators are similarly expressed in terms of irreducible tensors of $\grpsu{3}$.  Carrying out calculations in this framework often requires the coefficients of unitary transformation between coupled and uncoupled products of two irreps (coupling coefficients also known as Wigner or Clebsch-Gordan coefficients).  It also often requires coefficients of unitary transformations between products of three or more irreps coupled in different order (recoupling coefficients analogous to 6$j$ and 9$j$ symbols used in angular momentum recoupling).

A number of algorithms for calculating $\grpsu{3}$ coupling coefficients have been formulated~\cite{draayer,Quesne19872259,prakash,rowe3,Pan19985642,rowe4,Pan201670,dang2024}, and several codes calculating these coefficients have been developed~\cite{akiyama,kaeding95,kaeding96,draeger,bahri}. Among the most widely used is the Fortran library originally written by Akiyama and Draayer~\cite{akiyama}, which includes the coupling coefficients as well as the recoupling coefficients transforming between coupling orders ``(12)3'' and ``1(23)''. The Akiyama-Draayer (AD) code has since been augmented with several unpublished improvements and extended by Millener to include recoupling coefficients transforming between coupling orders ``(12)3'' and ``(13)2'' and recoupling coefficients for products of 4 irreps~\cite{millener}. However, the AD library has several limitations. It loses precision and can produce incorrect results when larger quantum numbers are involved, which limits, \textit{e.g.}, the model space and mass of nuclei in nuclear structure calculations. Moreover, it is written in an older form of the Fortran programming language, limiting optimization for present and future computer architectures.

In this paper, we present a library \texttt{ndsu3lib} for computing of $\grpsu{3}$ coupling coefficients for both the canonical and angular momentum group chains, as well as $\grpsu{3}$ recoupling coefficients for transforming between products of three or four irreps defined in different coupling order.  The Fortran library provides a fresh implementation of the original Draayer-Akiyama (DA) algorithms~\cite{draayer} and Millener's algorithms~\cite{millener}.  We furthermore explicate the principles and relations underlying the DA algorithm and document the implemented formulae.

The \texttt{ndsu3lib} library takes advantage of modern Fortran features to both extend the range of quantum numbers and improve computational speed and numerical accuracy for larger quantum numbers.  It is safe for OpenMP multithreaded computations and uses multiprecision arithmetic.  Wrappers are provided for easy integration with codes written in C and C++. The library is intended for use, among other applications, in symmetry guided \textit{ab initio} nuclear structure calculations, \textit{e.g.}, the symplectic no-core configuration interaction framework~\cite{irrepfamilies,mccoy2018:diss}.

Recently, in parallel with the development of the present library, a C++ implementation \texttt{SU3lib} of the DA algorithms has been developed by Dytrych \textit{et al.}~\cite{tdsu3lib}. This library similarly provides for OpenMP multithreaded operation and supports the use of multiprecision arithmetic.  Since the original AD code, \texttt{SU3lib}, and \texttt{ndsu3lib} are all based on the same underlying DA algorithm, they yield consistent sets of coupling coefficients, namely, following the same phase conventions and the same prescriptions for the resolution of inner and outer multiplicities.

We test the precision and performance of \texttt{ndsu3lib} and compare it to the AD library as well as to \texttt{SU3lib}. To evaluate the precision, we examine how well the computed coefficients obey the expected orthonormality relations for coupling and recoupling coefficients.  We find that our library works for a larger range of $\grpsu{3}$ quantum numbers and provides more accurate $\grpsu{3}\supset\grpso{3}$ coupling coefficients, which are of particular interest in nuclear physics, with large quantum numbers, than the AD library. Our library provides more accurate $\grpsu{3}\supset\grpu{1}\times\grpsu{2}$ coupling coefficients with large quantum numbers than the the AD library and \texttt{SU3lib}. For the recoupling coefficients, the precisions of the three libraries are similar. In our timing tests, the speeds of the three libraries are found to be comparable. 

In Sect.~\ref{sec:background} we define the adopted notation and present background information. In Sect.~\ref{sec:algorithm} we review the algorithms for $\grpsu{3}$ coupling and recoupling coefficients. In Sect.~\ref{sec:implementation} we describe the structure, implementation details, and usage of our library. In Sects.~\ref{sec:precision} and~\ref{sec:speed} we present validation and precision tests of our library as well as a study of its speed with comparison to the AD library and \texttt{SU3lib}.
 \section{Background}
\label{sec:background}

In physics applications involving $\grpsu{3}$, calculations are often carried out in a basis with definite $\grpsu{3}$ symmetry.  That is, the Hilbert space is decomposed into irreps of $\grpsu{3}$. An irrep of $\grpsu{3}$ can be further decomposed into irreps of the subgroups of $\grpsu{3}$.  Here we focus on subgroups commonly appearing in physics, namely, $\grpu{1}\times\grpsu{2}$ and $\grpso{3}$. In other words, we use a basis of the Hilbert space reducing either the canonical group chain $\grpsu{3}\supset\grpu{1}\times\grpsu{2}$ or the angular momentum group chain $\grpsu{3}\supset\grpso{3}$. The coupling coefficients for the canonical group chain are easy to compute, and then they can be transformed to the coupling coefficients for the angular momentum group chain.

We first define the bases of an $\grpsu{3}$ irrep which reduce either the canonical or angular momentum group chain (Sect.~\ref{sec:background:bases}). Then we define $\grpsu{3}$ coupling and the associated coupling coefficients and set up the outer multiplicity problem (Sect.~\ref{sec:background:coupling}). We also define the $\grpsu{3}$ coupling of $\grpsu{3}$ irreducible tensor operators, which is used in the formulation of the algorithm, and state the $\grpsu{3}$ Wigner-Eckart theorem (Sect.~\ref{sec:background:operators}).

\subsection{Bases of an irrep of $\grpsu{3}$}
\label{sec:background:bases}

Here we overview the bases of an $\grpsu{3}$ irrep which reduce either the canonical or angular momentum group chain and a relation between the two which will be used in Sect.~\ref{sec:algorithm:physical} describing computation of $\grpsu{3}\supset\grpso{3}$ coupling coefficients.

Following Elliott's convention~\cite{elliottI,elliottII}, an $\grpsu{3}$ irrep is labeled by the quantum numbers ($\lambda,\mu$). The quantum numbers labeling the states in a basis of the irrep $(\lambda,\mu)$ depend on the choice of group chain.

The basis states which reduce the canonical group chain are labeled by
\begin{equation}
\ketjh{(\lambda,\mu)}{\epsilon\Lambda M_{\Lambda}},
\end{equation}
where $\epsilon$ is the $\grpu{1}$ label, and $\Lambda$ is the $\grpsu{2}$ label, with $\grpsu{2}$ projection $M_{\Lambda}$:
\begin{equation}
\begin{array}{ccccccc}
\grpsu{3} & \supset & \grpu{1} & \times & \grpsu{2} & \supset & \grpu{1}. \\ (\lambda,\mu) & & \epsilon & & \Lambda & & M_{\Lambda} \\
\end{array}
\end{equation}
These quantum numbers are related to the hypercharge $Y$ and isospin $I$ used in particle physics: $\epsilon=-3Y$ and $\Lambda=I$~\cite{hecht-vcs}.

The possible values for $\epsilon$ and $\Lambda$ are given by the $\grpsu{3}$ to $\grpu{1}\times\grpsu{2}$ branching rule~\cite{hecht}:
\begin{align}
\epsilon&=2\lambda+\mu-3(p+q),\label{epsilon}\\ \Lambda&=\frac{\mu+p-q}{2},\label{Lambda}
\end{align}
where $p$ and $q$ are integers satisfying $0\le p\le\lambda$ and $0\le q\le\mu$.  The possible values of $M_\Lambda$ are given by the known angular momentum branching rule $M_{\Lambda}=-\Lambda,\dots,\Lambda$.

The basis states reducing the canonical group chain can be obtained by laddering from an extremal state with the $\grpsu{3}$ raising and lowering operators. The extremal state
\begin{equation}
\ketjh{(\lambda,\mu)}{\epsilon^{\rm E}\Lambda^{\rm E}M_{\Lambda}^{\rm E}}
\end{equation}
is either the highest-weight state
\begin{equation}
\ketjh{(\lambda,\mu)}{\epsilon^{\rm H}\Lambda^{\rm H}M_{\Lambda}^{\rm H}},
\end{equation}
which is annihilated by the $\grpsu{3}$ raising operators, or the lowest-weight state
\begin{equation}
\ketjh{(\lambda,\mu)}{\epsilon^{\rm L}\Lambda^{\rm L}M_{\Lambda}^{\rm L}},
\end{equation}
which is annihilated by the $\grpsu{3}$ lowering operators~\cite{hecht-vcs}. The highest-weight quantum numbers are given by
\begin{equation}
\epsilon^{\rm H}=-\lambda-2\mu,\quad\Lambda^{\rm H}=\frac{\lambda}{2},\quad M_{\Lambda}^{\rm H}=-\frac{\lambda}{2},
\end{equation}
and the lowest-weight quantum numbers are given by
\begin{equation}
\epsilon^{\rm L}=2\lambda+\mu,\quad\Lambda^{\rm L}=\frac{\mu}{2},\quad M_{\Lambda}^{\rm L}=\frac{\mu}{2}.
\end{equation}

The orthonormal basis states which reduce the angular momentum group chain are obtained by orthonormalization of the Elliott basis states~\cite{elliottII}. These Elliott basis states are obtained by projecting out states with good angular momentum from an extremal state~\cite{elliottII,draayerp}:
\begin{equation}\label{-proj}
\ketjh{(\lambda,\mu)}{KLM}=P^{L}_{MK}\ketjh{(\lambda,\mu)}{\epsilon^{\rm E}\Lambda^{\rm E}M_{\Lambda}^{\rm E}},
\end{equation}
where $L$ is the $\grpso{3}$ quantum number, \textit{i.e.}, the angular momentum with projection $M$ along the laboratory frame $z$-axis, and $K$ is the projection of $L$ along the body-fixed 3-axis. The quantum number $K$ here serves as an inner multiplicity index which distinguishes distinct $\grpso{3}$ irreps with the same quantum number $L$:
\begin{equation}
\begin{array}{ccccc}
\grpsu{3} & \supset & \grpso{3} & \supset & \grpso{2}. \\ (\lambda,\mu) & K & L & & M \\
\end{array}
\end{equation}
The possible values of $K$ and $L$ are given by~\cite{elliottI,elliottII,harvey1968:su3-shell}
\begin{align}\label{so3branch}
\nonumber K&=\min(\lambda,\mu),\,\min(\lambda,\mu)-2,\dots,1\,\,\textrm{or}\,\,0,\\ L&=\left\{\begin{array}{ll} K,K+1,\ldots,K+\max(\lambda,\mu), & \quad K\ne0,\\ \max(\lambda,\mu),\max(\lambda,\mu)-2,\ldots,1\,\,\textrm{or}\,\,0, & \quad K=0.\\
\end{array}\right.
\end{align}
The choice of the extremal state in the definition~(\ref{-proj}) is a matter of convention. In Elliott's convention~\cite{elliottII}, it depends on the values of $\lambda$ and $\mu$, in particular:
\begin{equation}\label{e}
\ketjh{(\lambda,\mu)}{\epsilon^{\rm E}\Lambda^{\rm E}M_{\Lambda}^{\rm E}}=\left\{\begin{array}{ll} \ketjh{(\lambda,\mu)}{\epsilon^{\rm H}\Lambda^{\rm H}M_{\Lambda}^{\rm H}}, & \quad\lambda<\mu, \\ \ketjh{(\lambda,\mu)}{\epsilon^{\rm L}\Lambda^{\rm L}M_{\Lambda}^{\rm L}}, & \quad\lambda\ge\mu.
\end{array}\right.
\end{equation}

The Elliott basis states are not normalized, nor are they orthogonal with respect to $K$. The orthonormal basis states are obtained by Gram-Schmidt orthonormalization of the Elliott basis states~\cite{vergados,draayer}:
\begin{equation}\label{ortos}
\ketjh{(\lambda,\mu)}{\kappa LM}=\sum_{j=1}^{\kappa}O^{(\lambda,\mu)L}_{\kappa j}\ketjh{(\lambda,\mu)}{K_{j}LM},
\end{equation}
where $\kappa=1,2,\ldots,\kappa_{\rm max}$ is simply a counting index, $K_{1},K_{2},\ldots,K_{\kappa_{\rm max}}$ are the possible values of $K$ for a given $L$ in ascending order, and $O^{(\lambda,\mu)L}$ is an orthonormalization matrix of size $\kappa_{\rm max}\times\kappa_{\rm max}$. Note that the orthonormal basis state in~(\ref{ortos}) is a linear combination of the Elliott basis states with $K_{j}$ where $j\le\kappa$, and thus with $K\leq K_\kappa$. Similarly like $K$, the index $\kappa$ serves as an inner multiplicity index distinguishing multiple occurrences of a given $L$ within the irrep $(\lambda,\mu)$. Thus, basis states which reduce the angular momentum group chain are labeled by
\begin{equation}
\begin{array}{ccccc}
\grpsu{3} & \supset & \grpso{3} & \supset & \grpso{2}. \\ (\lambda,\mu) & \kappa & L & & M \\
\end{array}
\end{equation}
Explicit formulae for $\kappa_{\rm max}$ and the possible values of $K$ for a given $L$ are given in~\ref{appendix:basis}, along with a recursive definition of the orthonormalization matrix $O^{(\lambda,\mu)L}$, given by~(\ref{6a})--(\ref{6c}).

The orthonormalization~(\ref{ortos}) allows us to obtain the transformation brackets between the orthonormal bases reducing the canonical and angular momentum group chains in terms of overlaps of the basis states reducing the canonical group chain and the Elliott basis states, for which an explicit formula~(\ref{26}) is known:
\begin{equation}\label{ortoc}
\overlapjh{(\lambda,\mu)}{\epsilon\Lambda M_{\Lambda}}{(\lambda,\mu)}{\kappa LM}=\sum_{j=1}^{\kappa}O^{(\lambda,\mu)L}_{\kappa j}\overlapjh{(\lambda,\mu)}{\epsilon\Lambda M_{\Lambda}}{(\lambda,\mu)}{K_{j}LM}.
\end{equation}

\subsection{$\grpsu{3}$ coupling and recoupling}
\label{sec:background:coupling}

The $\grpsu{3}$ coupling coefficients are coefficients of unitary transformation between coupled and uncoupled bases of irreps of $\grpsu{3}$: \begin{multline}\label{wcan}
\ketjh{(\lambda_{3},\mu_{3})}{\epsilon_{3}\Lambda_{3}M_{\Lambda_{3}}}_{\rho}\\
=\sum_{\substack{\epsilon_{1}\Lambda_{1}M_{\Lambda_{1}}\Lambda_{2}\\(\epsilon_{2}M_{\Lambda_{2}})}}\wig{(\lambda_{1},\mu_{1})}{\epsilon_{1}\Lambda_{1}M_{\Lambda_{1}}}{(\lambda_{2},\mu_{2})}{\epsilon_{2}\Lambda_{2}M_{\Lambda_{2}}}{(\lambda_{3},\mu_{3})}{\epsilon_{3}\Lambda_{3}M_{\Lambda_{3}}}_{\rho}\\
\times\ketjh{(\lambda_{1},\mu_{1})}{\epsilon_{1}\Lambda_{1}M_{\Lambda_{1}}}\ketjh{(\lambda_{2},\mu_{2})}{\epsilon_{2}\Lambda_{2}M_{\Lambda_{2}}},
\end{multline}
where the transformation coefficients are $\grpsu{3}\supset\grpu{1}\times\grpsu{2}$ coupling coefficients. Note that the quantum numbers $\epsilon$ and $M_{\Lambda}$ are additive, \textit{i.e.}, $\epsilon_{1}+\epsilon_{2}=\epsilon_{3}$ and $M_{\Lambda_{1}}+M_{\Lambda_{2}}=M_{\Lambda_{3}}$. This constrains the sum in~(\ref{wcan}), which effectively reduces to a summation over only $\epsilon_{1}$, $\Lambda_{1}$, $\Lambda_{2}$, and $M_{\Lambda_{1}}$.  The remaining, redundant summation indices in~(\ref{wcan}) are shown in parentheses.

In the product space, there can be multiple linearly independent irrepsof $\grpsu{3}$ which each separately transform under $\grpsu{3}$ as the $(\lambda_3,\mu_3)$ irrep. The label $\rho$ distinguishes between these irreps, with bases given by~(\ref{wcan}). Together these irreps form a larger space characterized by the same definite symmetry labels $(\lambda_3,\mu_3)$. However, the separation according to $\rho$ is arbitrary~\cite{butler}. It is readily verified that states formed as an arbitrary linear combination of the bases for these irreps again form the basis for an irrep of $\grpsu{3}$, transforming as $(\lambda_3,\mu_3)$. Thus, in the presence of an outer multiplicity ($\rho=1,2,\ldots,\rho_{\rm max}$), the orthonormal set of coupled states is only defined to within an arbitrary unitary transformation. Namely, ``primed'' and ``unprimed'' orthonormal sets of coupled states are related by
\begin{equation}
\ketjh{(\lambda_3,\mu_3)}{\epsilon_3\Lambda_3 M_{\Lambda_3}}_{\rho'}'=\sum_{\rho}A_{\rho'\rho}\ketjh{(\lambda_3,\mu_3)}{\epsilon_3\Lambda_3 M_{\Lambda_3}}_{\rho},
\end{equation}
where $A$ is a unitary matrix.

Rewritten in terms of coupling coefficients, this ambiguity in choice of basis for the coupled space is reflected in the existence of alternative valid choices of orthonormal sets of coupling coefficients. Such ``primed'' and ``unprimed'' coupling coefficients are similarly related by a unitary transformation as
\begin{multline}\label{eqn:cg-unitary-xform}
\wig{(\lambda_{1},\mu_{1})}{\epsilon_{1}\Lambda_{1}M_{\Lambda_{1}}}{(\lambda_{2},\mu_{2})}{\epsilon_{2}\Lambda_{2}M_{\Lambda_{2}}}{(\lambda_{3},\mu_{3})}{\epsilon_{3}\Lambda_{3}M_{\Lambda_{3}}}_{\rho'}'\\
=\sum_{\rho}A_{\rho'\rho}\wig{(\lambda_{1},\mu_{1})}{\epsilon_{1}\Lambda_{1}M_{\Lambda_{1}}}{(\lambda_{2},\mu_{2})}{\epsilon_{2}\Lambda_{2}M_{\Lambda_{2}}}{(\lambda_{3},\mu_{3})}{\epsilon_{3}\Lambda_{3}M_{\Lambda_{3}}}_{\rho}.
\end{multline}
In fact, in the DA algorithm (Sect.~\ref{sec:algorithm}), we shall have reason to consider, as an intermediate result, a non-orthonormal set of primed states, and thus a non-orthonormal set of coupling coefficients, in which case the transformation coefficients $A_{\rho'\rho}$ no longer constitute a unitary matrix.

The problem of choosing a particular basis for the coupled space, and thus the meaning of the outer multiplicity label, is known as the ``outer multiplicity problem''. While any resolution of the outer multiplicity problem yields a valid set of coupling coefficients, for consistency between calculations, it is essential that an algorithm for generating coupling coefficients provide a replicable resolution of the outer multiplicity.

Moreover, some choices may be more convenient than others. For $\grpsu{3}$, the canonical solution to the outer multiplicity problem is provided by the Biedenharn-Louck-Hecht (BLH) prescription~\cite{hecht,bl1,bl2,bl3,castilhoalcaras1970:su3-outer-multiplicity-27plet,asherova1997:u3-outer-multiplicity}. This prescription is formally motivated in terms of null space properties of Wigner operators, which are $\grpsu{3}$ irreducible tensor operators, the matrix elements of which define the coupling coefficients. Numerically, the BLH prescription may be imposed by requiring coupling coefficients which satisfy a certain condition [given by~(\ref{blh}) below] to vanish. Further discussion may be found in Refs.~\cite{rowe4,bahri}.

In the angular momentum reduction scheme, a basis state in a coupled irrep is given by
\begin{multline}\label{wphys}
\ketjh{(\lambda_{3},\mu_{3})}{\kappa_{3}L_{3}M_{3}}_{\rho}\\
=\sum_{\substack{\kappa_{1}L_{1}M_{1}\\\kappa_{2}L_{2}(M_{2})}}\wig{(\lambda_{1},\mu_{1})}{\kappa_{1}L_{1}M_{1}}{(\lambda_{2},\mu_{2})}{\kappa_{2}L_{2}M_{2}}{(\lambda_{3},\mu_{3})}{\kappa_{3}L_{3}M_{3}}_{\rho}\\
\times\ketjh{(\lambda_{1},\mu_{1})}{\kappa_{1}L_{1}M_{1}}\ketjh{(\lambda_{2},\mu_{2})}{\kappa_{2}L_{2}M_{2}},
\end{multline}
where the transformation coefficients are $\grpsu{3}\supset\grpso{3}$ coupling coefficients. Note that the quantum number $M$ is additive, \textit{i.e.}, $M_{1}+M_{2}=M_{3}$, which constrains the summation in~(\ref{wphys}).

An $\grpsu{3}$ coupling coefficient can be factored into a reduced coupling coefficient (RCC) independent of the projections $M_{\Lambda}$ or $M$ and an $\grpsu{2}$ or $\grpso{3}$ coupling coefficient, which carries all the dependence on the projections. The RCC is indicated by a double bar in the following expressions:
\begin{multline}\label{racfac}
\wig{(\lambda_{1},\mu_{1})}{\epsilon_{1}\Lambda_{1}M_{\Lambda_{1}}}{(\lambda_{2},\mu_{2})}{\epsilon_{2}\Lambda_{2}M_{\Lambda_{2}}}{(\lambda_{3},\mu_{3})}{\epsilon_{3}\Lambda_{3}M_{\Lambda_{3}}}_{\rho}\\\!=\!\rwig{(\lambda_{1},\mu_{1})}{\epsilon_{1}\Lambda_{1}}{(\lambda_{2},\mu_{2})}{\epsilon_{2}\Lambda_{2}}{(\lambda_{3},\mu_{3})}{\epsilon_{3}\Lambda_{3}}_{\rho}\!\wig{\Lambda_{1}}{\!M_{\Lambda_{1}}}{\Lambda_{2}}{M_{\Lambda_{2}}\!}{\Lambda_{3}}{\!M_{\Lambda_{3}}\!}\!
\end{multline}
and
\begin{multline}\label{eqn:factorization:physical}
\wig{(\lambda_{1},\mu_{1})}{\kappa_{1}L_{1}M_{1}}{(\lambda_{2},\mu_{2})}{\kappa_{2}L_{2}M_{2}}{(\lambda_{3},\mu_{3})}{\kappa_{3}L_{3}M_{3}}_{\rho}\\=\rwig{(\lambda_{1},\mu_{1})}{\kappa_{1}L_{1}}{(\lambda_{2},\mu_{2})}{\kappa_{2}L_{2}}{(\lambda_{3},\mu_{3})}{\kappa_{3}L_{3}}_{\rho}\wig{L_{1}}{M_{1}}{L_{2}}{M_{2}}{L_{3}}{M_{3}}.
\end{multline}
Since the $\grpsu{2}$ (Eq.~\ref{racfac}) or $\grpso{3}$ (Eq.~\ref{eqn:factorization:physical}) coupling coefficients are readily available, the problem of obtaining the $\grpsu{3}$ coupling coefficients reduces to that of obtaining the RCCs.

The RCCs obey the orthonormality relations
\begin{align}
\nonumber\sum_{(\epsilon_{1})\Lambda_{1}\epsilon_{2}\Lambda_{2}}&\rwig{(\lambda_{1},\mu_{1})}{\epsilon_{1}\Lambda_{1}}{(\lambda_{2},\mu_{2})}{\epsilon_{2}\Lambda_{2}}{(\lambda_{3},\mu_{3})}{\epsilon_{3}\Lambda_{3}}_{\rho}\\ \times&\rwig{(\lambda_{1},\mu_{1})}{\epsilon_{1}\Lambda_{1}}{(\lambda_{2},\mu_{2})}{\epsilon_{2}\Lambda_{2}}{(\lambda_{3},\mu_{3})}{\epsilon_{3}\Lambda_{3}}_{\rho'}=\delta_{\rho\rho'},\label{ortocan}\\ \nonumber\sum_{\kappa_{1}L_{1}\kappa_{2}L_{2}}&\rwig{(\lambda_{1},\mu_{1})}{\kappa_{1}L_{1}}{(\lambda_{2},\mu_{2})}{\kappa_{2}L_{2}}{(\lambda_{3},\mu_{3})}{\kappa_{3}L_{3}}_{\rho}\\ \times&\rwig{(\lambda_{1},\mu_{1})}{\kappa_{1}L_{1}}{(\lambda_{2},\mu_{2})}{\kappa_{2}L_{2}}{(\lambda_{3},\mu_{3})}{\kappa_{3}L_{3}}_{\rho'}=\delta_{\rho\rho'}.\label{ortowp}
\end{align}

When more than two $\grpsu{3}$ irreps need to be coupled, the resulting $\grpsu{3}$ irrep can be constructed in different ways depending on the order of the coupling.

Transformations between different orders of coupling of three $\grpsu{3}$ irreps involve the $U$~\cite{hecht,hecht2} and $Z$~\cite{millener} recoupling coefficients (unitary 6-($\lambda,\mu$) coefficients analogous to the 6$j$ symbols known from the angular momentum recoupling). In particular, the ``(12)3'' coupling $[[(\lambda_1,\mu_1)\times(\lambda_2,\mu_2)]\times(\lambda_3,\mu_3)]$ is related to the ``1(23)'' coupling $[(\lambda_1,\mu_1)\times[(\lambda_2,\mu_2)\times(\lambda_3,\mu_3)]]$ via the $U$ coefficients and to the ``(13)2'' coupling $[[(\lambda_1,\mu_1)\times(\lambda_3,\mu_3)]\times(\lambda_2,\mu_2)]$ via the $Z$ coefficients. Transformations between different orders of coupling of four $\grpsu{3}$ irreps involve the 9-($\lambda,\mu$) coefficients~\cite{hecht9lm,hecht2,millener} (analogous to the 9$j$ symbols known from the angular momentum recoupling). In particular, these are the coefficients of the transformation between the ``(12)(34)'' coupling $[[(\lambda_1,\mu_1)\times(\lambda_2,\mu_2)]\times[(\lambda_3,\mu_3)\times(\lambda_4,\mu_4)]]$ and the ``(13)(24)'' coupling $[[(\lambda_1,\mu_1)\times(\lambda_3,\mu_3)]\times[(\lambda_2,\mu_2)\times(\lambda_4,\mu_4)]]$.

\subsection{$\grpsu{3}$ irreducible tensor operators}
\label{sec:background:operators}

Now we define $\grpsu{3}$ coupling of $\grpsu{3}$ irreducible tensor operators. An $\grpsu{3}$ irreducible tensor operator $T^{(\lambda,\mu)}$ is a tensor operator transforming with respect to the group $\grpsu{3}$ according to the irrep ($\lambda,\mu$). Two $\grpsu{3}$ irreducible tensor operators $T^{(\lambda_{1},\mu_{1})}$ and $T^{(\lambda_{2},\mu_{2})}$ can be coupled to yield, as their product, an $\grpsu{3}$ irreducible tensor operator $\left[T^{(\lambda_{1},\mu_{1})}\times T^{(\lambda_{2},\mu_{2})}\right]^{\rho(\lambda_{3},\mu_{3})}$. In the $\grpsu{3}\supset\grpu{1}\times\grpsu{2}$ scheme the components of this operator are
\begin{multline}\label{eqn:coupled-product-tensor}
\left[T^{(\lambda_{1},\mu_{1})}\times T^{(\lambda_{2},\mu_{2})}\right]^{\rho(\lambda_{3},\mu_{3})}_{\epsilon_{3}\Lambda_{3}M_{\Lambda_{3}}}\\
=\sum_{\substack{\epsilon_{1}\Lambda_{1}M_{\Lambda_{1}}\Lambda_{2}\\(\epsilon_2M_{\Lambda_{2}})}}
\wig{(\lambda_{1},\mu_{1})}{\epsilon_{1}\Lambda_{1}M_{\Lambda_{1}}}{(\lambda_{2},\mu_{2})}{\epsilon_{2}\Lambda_{2}M_{\Lambda_{2}}}{(\lambda_{3},\mu_{3})}{\epsilon_{3}\Lambda_{3}M_{\Lambda_{3}}}_{\rho}\\
\times T^{(\lambda_{1},\mu_{1})}_{\epsilon_{1}\Lambda_{1}M_{\Lambda_{1}}}T^{(\lambda_{2},\mu_{2})}_{\epsilon_{2}\Lambda_{2}M_{\Lambda_{2}}}.
\end{multline}

According to the Wigner-Eckart theorem for $\grpsu{3}\supset\grpu{1}\times\grpsu{2}$, the $\grpsu{2}$-reduced matrix elements (RMEs) of an $\grpsu{3}$ irreducible tensor operator $T^{(\lambda_{2},\mu_{2})}$ can be expressed in terms of matrix elements $\langle(\lambda_{3},\mu_{3})||T^{(\lambda_{2},\mu_{2})}||(\lambda_{1},\mu_{1})\rangle_{\rho}$ furthermore reduced with respect to $\grpsu{3}$, as
\begin{multline}\label{9}
\rbrajh{(\lambda_{3},\mu_{3})}{\epsilon_{3}\Lambda_{3}}T_{\epsilon_{2}\Lambda_{2}}^{(\lambda_{2},\mu_{2})}\rketjh{(\lambda_{1},\mu_{1})}{\epsilon_{1}\Lambda_{1}}\\
=\sum_{\rho}\langle(\lambda_{3},\mu_{3})||T^{(\lambda_{2},\mu_{2})}||(\lambda_{1},\mu_{1})\rangle_{\rho}\\
\times{\rwig{(\lambda_{1},\mu_{1})}{\epsilon_{1}\Lambda_{1}}{(\lambda_{2},\mu_{2})}{\epsilon_{2}\Lambda_{2}}{(\lambda_{3},\mu_{3})}{\epsilon_{3}\Lambda_{3}}}_{\rho}.
\end{multline}
Note that this Wigner-Eckart theorem for $\grpsu{3}$ involves a sum over the outer multiplicity index $\rho$, which is not present in the Wigner-Eckart theorem for the simpler case of $\grpsu{2}$.  (An analogous Wigner-Eckart theorem may be written for the $\grpsu{3}\supset\grpso{3}$ scheme, but it is not needed in the following discussions.)

 \section{Algorithms}
\label{sec:algorithm}

We review the DA and Millener's algorithms, implemented in \texttt{ndsu3lib}, for calculation of $\grpsu{3}$ RCCs for the canonical (Sect.~\ref{sec:algorithm:canonical}) and angular momentum (Sect.~\ref{sec:algorithm:physical}) group chains, and for calculation of recoupling coefficients (Sect.~\ref{sec:algorithm:recoupling}).

\subsection{$\grpsu{3}\supset\grpu{1}\times\grpsu{2}$ reduced coupling coefficients}
\label{sec:algorithm:canonical}
 
The DA algorithm provides a scheme for calculating $\grpsu{3}\supset\grpu{1}\times\grpsu{2}$ RCCs which, moreover, are constructed so as to satisfy the BLH prescription (Sect.~\ref{sec:background:coupling}) for resolving the outer multiplicity problem.

The algorithm makes use of the fundamental recurrence relations~\cite{hecht} connecting different coupling coefficients for the same coupling $(\lambda_1,\mu_1)\times(\lambda_2,\mu_2)\to(\lambda_3,\mu_3)$, obtained by the method of infinitesimal generators, that is, by considering the laddering action of the group generators within these irreps. However, these recurrence relations apply equally well to any valid set of coupling coefficients, and do not, in themselves, resolve the outer multiplicity problem.

The DA algorithm furthermore ensures that the calculated coupling coefficients satisfy the BLH prescription. It does so through a particular choice of seed coefficients for the recurrence stemming from the method of infinitesimal generators. These seeds are generated by relating the RCCs for the given coupling $(\lambda_1,\mu_1)\times(\lambda_2,\mu_2)\to(\lambda_3,\mu_3)$ to simpler RCCs arising for couplings $(\lambda_1,\mu_1)\times(\bar{\lambda}_2,\bar{\mu}_2)\to(\lambda_3,\mu_3)$, with $\bar{\lambda}_2<\lambda_2$ and $\bar{\mu}_2<\mu_2$, through a building-up process.  This building-up process is derived by relating the RCCs to matrix elements of a suitably defined Wigner operator, devised such that the resulting RCCs are guaranteed to satisfy the BLH prescription by construction.

To elucidate the DA algorithm, as implemented in the present code, we first review the standard recurrence relations (Sect.~\ref{sec:algorithm:canonical:infinitesimal-generators}), then detail how the DA algorithm ensures that the calculated coupling coefficients satisfy the BLH prescription (Sect.~\ref{sec:algorithm:canonical:building-up}), then put these ideas together to see how they determine the recurrence scheme for the RCCs  (Sect.~\ref{sec:algorithm:canonical:da}).  We focus here on the principal ideas and equations, defering some details to~\ref{appendix}.

\subsubsection{Method of infinitesimal generators}
\label{sec:algorithm:canonical:infinitesimal-generators}

The method of infinitesimal generators provides relations between different RCCs for the same coupling $(\lambda_1,\mu_1)\times(\lambda_2,\mu_2)\to(\lambda_3,\mu_3)$ (given for an arbitrary subgroup chain by (11) or (A7) of Ref.~\cite{racah}), by considering the action of the same generator, acting either on an uncoupled product state or a coupled product state, and relating the two results. This approach was notably applied by Racah~\cite{racahmethod} and is thus also known as “Racah's method”~\cite{wybourne}.  It was applied to $\grpsu{3}\supset\grpu{1}\times\grpsu{2}$ coupling coefficients by Hecht~\cite{hecht}.

Let us decompose the generators of $\grpsu{3}$ [which transform as the adjoint irrep $(1,1)$] into a set of tensors $C^{(1,1)}_{\epsilon_T\Lambda_T}$ with respect to $\grpu{1}\times\grpsu{2}$ as well. Then the relations between $\grpsu{3}\supset\grpu{1}\times\grpsu{2}$ RCCs provided by the method of infinitesimal generators are of the form
\begin{multline}\label{racgen}
\rwig{(\lambda_{1},\mu_{1})}{\epsilon_{1}\Lambda_{1}}{(\lambda_{2},\mu_{2})}{\epsilon_{2}\Lambda_{2}}{(\lambda_{3},\mu_{3})}{\epsilon_{3}+\epsilon_T,\Lambda_{3}}_{\rho}\\
=\rbrajh{(\lambda_{3},\mu_{3})}{\epsilon_{3}+\epsilon_T,\Lambda_{3}}C^{(1,1)}_{\epsilon_T\Lambda_T}\rketjh{(\lambda_{3},\mu_{3})}{\epsilon_{3}\Lambda_{3}'}^{-1}\\
\times\Bigg[\sum_{\Lambda_{1}'}(-1)^{\Lambda_{3}-\Lambda_{1}-\Lambda_{2}}(-1)^{\Lambda_{3}'-\Lambda_{1}'-\Lambda_{2}}\\
\times U\left(\Lambda_2\Lambda_1'\Lambda_3\Lambda_T;\Lambda_3'\Lambda_1\right)\\
\times\rbrajh{(\lambda_{1},\mu_{1})}{\epsilon_{1}\Lambda_{1}}C^{(1,1)}_{\epsilon_T\Lambda_T}\rketjh{(\lambda_{1},\mu_{1})}{\epsilon_{1}-\epsilon_T,\Lambda_{1}'}\\
\times\rwig{(\lambda_{1},\mu_{1})}{\epsilon_{1}-\epsilon_T,\Lambda_{1}'}{(\lambda_{2},\mu_{2})}{\epsilon_{2}\Lambda_{2}}{(\lambda_{3},\mu_{3})}{\epsilon_{3}\Lambda_{3}'}_{\rho}\\
+\sum_{\Lambda_{2}'}U\left(\Lambda_1\Lambda_2'\Lambda_3\Lambda_T;\Lambda_3'\Lambda_2\right)\\
\times\rbrajh{(\lambda_{2},\mu_{2})}{\epsilon_{2}\Lambda_{2}}C^{(1,1)}_{\epsilon_T\Lambda_T}\rketjh{(\lambda_{2},\mu_{2})}{\epsilon_{2}-\epsilon_T,\Lambda_{2}'}\\
\times\rwig{(\lambda_{1},\mu_{1})}{\epsilon_{1}\Lambda_{1}}{(\lambda_{2},\mu_{2})}{\epsilon_{2}-\epsilon_T,\Lambda_{2}'}{(\lambda_{3},\mu_{3})}{\epsilon_{3}\Lambda_{3}'}_{\rho}\Bigg].
\end{multline}
Here the RCCs are obtained in terms of generator RMEs [reduced with respect to $\grpsu{2}$] and unitary recoupling coefficients $U$ for $\grpsu{2}$.

Note that the relations~(\ref{racgen}) are linear, homogeneous relations among multiplets of RCCs, sharing the same $(\lambda,\mu)$ quantum numbers but differing in the $\epsilon\Lambda$ quantum numbers. The only useful relations are obtained by considering, from among the $\grpsu{3}$ generators $C^{(1,1)}_{\epsilon_T\Lambda_T}$, the $\epsilon$-raising generator $C^{(1,1)}_{+3,1/2}$ and the $\epsilon$-lowering generator $C^{(1,1)}_{-3,1/2}$. [That is, we exclude the $\grpu{1}$ generator $C^{(1,1)}_{00}$ and $\grpsu{2}$ generator $C^{(1,1)}_{01}$. Otherwise, the relations fail to connect RCCs involving different $\epsilon\Lambda$ labels.] The resulting relations are\footnote{Relation~(\ref{19}) corresponds to~(19) of Ref.~\cite{draayer}, where, in the last row, $q_i$ should be $p_i$.}
\begin{multline}\label{19}
\rwig{(\lambda_{1},\mu_{1})}{\epsilon_{1}\Lambda_{1}}{(\lambda_{2},\mu_{2})}{\epsilon_{2}\Lambda_{2}}{(\lambda_{3},\mu_{3})}{\epsilon_{3}+3,\Lambda_{3}}_{\rho}\\
=\rbrajh{(\lambda_{3},\mu_{3})}{\epsilon_{3}+3,\Lambda_{3}}C^{(1,1)}_{+3,\frac{1}{2}}\rketjh{(\lambda_{3},\mu_{3})}{\epsilon_{3}\Lambda_{3}'}^{-1}\\
\times\Bigg[\sum_{\Lambda_{1}'=\Lambda_1\pm\frac{1}{2}}(-1)^{\Lambda_{3}-\Lambda_{1}-\Lambda_{2}}(-1)^{\Lambda_{3}'-\Lambda_{1}'-\Lambda_{2}}\\
\times U\left(\Lambda_2\Lambda_1'\Lambda_3\frac{1}{2};\Lambda_3'\Lambda_1\right)\\
\times\rbrajh{(\lambda_{1},\mu_{1})}{\epsilon_{1}\Lambda_{1}}C^{(1,1)}_{+3,\frac{1}{2}}\rketjh{(\lambda_{1},\mu_{1})}{\epsilon_{1}-3,\Lambda_{1}'}\\
\times\rwig{(\lambda_{1},\mu_{1})}{\epsilon_{1}-3,\Lambda_{1}'}{(\lambda_{2},\mu_{2})}{\epsilon_{2}\Lambda_{2}}{(\lambda_{3},\mu_{3})}{\epsilon_{3}\Lambda_{3}'}_{\rho}\\
+\sum_{\Lambda_{2}'=\Lambda_2\pm\frac{1}{2}}U\left(\Lambda_1\Lambda_2'\Lambda_3\frac{1}{2};\Lambda_3'\Lambda_2\right)\\
\times\rbrajh{(\lambda_{2},\mu_{2})}{\epsilon_{2}\Lambda_{2}}C^{(1,1)}_{+3,\frac{1}{2}}\rketjh{(\lambda_{2},\mu_{2})}{\epsilon_{2}-3,\Lambda_{2}'}\\
\times\rwig{(\lambda_{1},\mu_{1})}{\epsilon_{1}\Lambda_{1}}{(\lambda_{2},\mu_{2})}{\epsilon_{2}-3,\Lambda_{2}'}{(\lambda_{3},\mu_{3})}{\epsilon_{3}\Lambda_{3}'}_{\rho}\Bigg],
\end{multline}
where $\Lambda_3'=\Lambda_3\pm\frac{1}{2}$, and
\begin{multline}\label{19conjug}
\rwig{(\lambda_{1},\mu_{1})}{\epsilon_{1}\Lambda_{1}}{(\lambda_{2},\mu_{2})}{\epsilon_{2}\Lambda_{2}}{(\lambda_{3},\mu_{3})}{\epsilon_{3}-3,\Lambda_{3}}_{\rho}\\
=\rbrajh{(\lambda_{3},\mu_{3})}{\epsilon_{3}-3,\Lambda_{3}}C^{(1,1)}_{-3,\frac{1}{2}}\rketjh{(\lambda_{3},\mu_{3})}{\epsilon_{3}\Lambda_{3}'}^{-1}\\
\times\Bigg[\sum_{\Lambda_{1}'=\Lambda_1\pm\frac{1}{2}}(-1)^{\Lambda_{3}-\Lambda_{1}-\Lambda_{2}}(-1)^{\Lambda_{3}'-\Lambda_{1}'-\Lambda_{2}}\\
\times U\left(\Lambda_2\Lambda_1'\Lambda_3\frac{1}{2};\Lambda_3'\Lambda_1\right)\\
\times\rbrajh{(\lambda_{1},\mu_{1})}{\epsilon_{1}\Lambda_{1}}C^{(1,1)}_{-3,\frac{1}{2}}\rketjh{(\lambda_{1},\mu_{1})}{\epsilon_{1}+3,\Lambda_{1}'}\\
\times\rwig{(\lambda_{1},\mu_{1})}{\epsilon_{1}+3,\Lambda_{1}'}{(\lambda_{2},\mu_{2})}{\epsilon_{2}\Lambda_{2}}{(\lambda_{3},\mu_{3})}{\epsilon_{3}\Lambda_{3}'}_{\rho}\\
+\sum_{\Lambda_{2}'=\Lambda_2\pm\frac{1}{2}}U\left(\Lambda_1\Lambda_2'\Lambda_3\frac{1}{2};\Lambda_3'\Lambda_2\right)\\
\times\rbrajh{(\lambda_{2},\mu_{2})}{\epsilon_{2}\Lambda_{2}}C^{(1,1)}_{-3,\frac{1}{2}}\rketjh{(\lambda_{2},\mu_{2})}{\epsilon_{2}+3,\Lambda_{2}'}\\
\times\rwig{(\lambda_{1},\mu_{1})}{\epsilon_{1}\Lambda_{1}}{(\lambda_{2},\mu_{2})}{\epsilon_{2}+3,\Lambda_{2}'}{(\lambda_{3},\mu_{3})}{\epsilon_{3}\Lambda_{3}'}_{\rho}\Bigg],
\end{multline}
respectively. In~(\ref{19}) and~(\ref{19conjug}) the RCCs are obtained in terms of the generator RMEs, which are available, \textit{e.g.}, in~\cite{hecht,hecht-vcs}, and $\grpsu{2}$ recoupling coefficients of a class for which explicit expressions are available, \textit{e.g.}, in~\cite{var}.

These relations apply to any valid set of RCCs, independent of the outer multiplicity index $\rho$.  Such a set of linear, homogeneous relations does not define the overall phase of the resulting RCCs, nor does it impose their orthonormality with respect to the outer multiplicity index $\rho$. Orthonormality must be imposed, independently, by imposing orthonormality of the coupled states, which implies~(\ref{ortocan}).

\subsubsection{Building-up process}
\label{sec:algorithm:canonical:building-up}

The BLH prescription for the resolution of the outer multiplicity is formally motivated in terms of null space properties of Wigner operators~\cite{bl1,bl2,bl3}, which are $\grpsu{3}$ irreducible tensor operators, the matrix elements of which define the coupling coefficients. However, numerically, the BLH prescription may be imposed by requiring certain RCCs to vanish.  Namely,
\begin{multline}
  \label{blh}
  \rwig{(\lambda_1,\mu_1)}{\epsilon_1\Lambda_1}{(\lambda_2,\mu_2)}{\epsilon_2\Lambda_2}{(\lambda_3,\mu_3)}{\epsilon_3\Lambda_3}_{\rho}=0\\
\text{for~}  |\Lambda_1-\Lambda_3|>\frac{1}{2}(\lambda_2+\mu_2-\eta_{\rm max}+\rho),
\end{multline}
where $\eta_{\rm max}$ is the positive integer such that the coupling $(\lambda_{1},\mu_{1})\times(\lambda_{2}-\eta_{\rm max}+1,\mu_{2}-\eta_{\rm max}+1)\to(\lambda_{3},\mu_{3})$ has unit multiplicity, while the coupling $(\lambda_{1},\mu_{1})\times(\lambda_{2}-\eta_{\rm max},\mu_{2}-\eta_{\rm max})\to(\lambda_{3},\mu_{3})$ is not allowed.  Note that the number of RCCs vanishing according to~(\ref{blh}) decreases with increasing $\rho$, and, in the case where $\eta_{\mathrm{max}}=\rho_{\mathrm{max}}$ (in general, $\eta_{\mathrm{max}}\geq\rho_{\mathrm{max}}$~\cite{draayer}), no vanishings are imposed by~(\ref{blh}) among the set of RCCs for maximal $\rho$, beyond those already implied by the $\grpsu{2}$ triangle inequality.

Given a complete set of RCCs for the coupling $(\lambda_{1},\mu_{1})\times(\lambda_{2},\mu_{2})\to(\lambda_{3},\mu_{3})$, \textit{e.g.}, obtained by the method of infinitesimal generators, we could construct from these a set of RCCs satisfying the BLH prescription simply by applying an appropriate unitary transformation~(\ref{eqn:cg-unitary-xform}).\footnote{If the sets of RCCs for different $\rho$ are arranged as row vectors, and segmented into blocks representing groups of RCCs sharing the same subgroup labels $\epsilon_3 \Lambda_3$ for the coupled state, analogously to Fig.~5(c) of Ref.~\cite{racah}, the BLH vanishing condition~(\ref{blh}) may be interpreted (assuming an appropriate ordering of the RCCs) as imposing an upper triangular pattern of zeros within certain blocks.  The unitary transformation to obtain such RCCs is therefore straightforward to determine, \textit{e.g.}, by row reduction followed by Gram-Schmidt orthonormalization.}  However, in practice, it is desirable to be able to selectively evaluate targeted subsets of RCCs, while still ensuring that they correspond to the BLH resolution of the outer multiplicity.  In particular, we shall see that so-called ``extremal'' RCCs, those with $\epsilon_3 \Lambda_3$ of highest or lowest weight, are of special relevance in evaluating the RCCs for the angular momentum chain (Sect.~\ref{sec:algorithm:physical}) and in evaluating recoupling coefficients (Sect.~\ref{sec:algorithm:recoupling}).

In the DA algorithm, a direct path to these extremal RCCs~--- and, crucially, one which by construction enforces the BLH prescription~--- is provided through a building-up process~\cite{castilhoalcaras1970:su3-outer-multiplicity-27plet,draayer}, which allows RCCs for the coupling $(\lambda_{1},\mu_{1})\times(\lambda_{2},\mu_{2})\to(\lambda_{3},\mu_{3})$ to be obtained recursively from RCCs for couplings $(\lambda_{1},\mu_{1})\times(\bar{\lambda}_{2},\bar{\mu}_{2})\to(\lambda_{3},\mu_{3})$ having lower outer multiplicity.  If non-extremal RCCs are sought, these may be found from the extremal RCCs thus obtained, by subsequent application of the recurrence relations from the method of infinitesimal generators, within the coupling $(\lambda_{1},\mu_{1})\times(\lambda_{2},\mu_{2})\to(\lambda_{3},\mu_{3})$.

Specifically, the recurrence is carried out separately for each value of the outer multiplicity index ($\rho=1,\ldots,\rho_{\mathrm{max}}$), chosed to start in each case from RCCs for the coupling with $(\bar\lambda_{2},\bar\mu_{2}) \equiv (\lambda_{2}-\eta,\mu_{2}-\eta)$, where $\eta\equiv \eta_{\mathrm{max}}-\rho$.\footnote{Note that $\rho$ is indeed a valid outer multiplicity index for the coupling $(\lambda_{1},\mu_{1})\times(\bar\lambda_{2},\bar\mu_{2})\to(\lambda_{3},\mu_{3})$, in fact, the maximal outer multiplicity index for this coupling~\cite{draayer}.}  Moreover, in deriving the recurrence, it will be helpful to keep in mind that we only need to obtain a valid set of RCCs, satisfying the BLH vanishing conditions~(\ref{blh}), for each specific value of the outer multiplicity index $\rho$, without concern for orthogonality of the sets of RCCs for different $\rho$ or, indeed, overall normalization for any given set.  These conditions may be imposed later by a Gram-Schmidt orthonormalization, with respect to the outer multiplicity index.  However, such Gram-Schmidt orthonormalization must be performed in order of \textit{increasing} $\rho$ (that is, decreasing number of enforced zeros), in order to preserve the BLH vanishing conditions.

The recurrence relation for a bulding-up process~\cite{hecht1967:so5-shell-wigner,hecht1969:su4-wigner} can, in general, be deduced simply from the summation identities relating a ``$(12)3$--$1(23)$'' recoupling coefficient to sums of products of RCCs [see~(\ref{22}) below].  The resulting relation is given for an arbitrary subgroup chain in~(19.207) of Ref.~\cite{wybourne}.  However, the DA algorithm makes use of a special form of such a building-up relation, one which enforces the BLH prescription, deduced by introducing an auxiliary operator and relating the RCCs to RMEs of this operator.

First, we define the Wigner operators $K^{(\lambda_{2},\mu_{2})\rho}$ ($\rho=1,\ldots,\rho_{\mathrm{max}}$), acting between the representation spaces for $(\lambda_{3},\mu_{3})$ and $(\lambda_{1},\mu_{1})$.  Each has just a single nonvanishing (and unit) $\grpsu{3}$-RME
\begin{equation}\label{eqn:wigner-su3-rme}
  \langle(\lambda_{3},\mu_{3})||K^{(\lambda_{2},\mu_{2})\rho}||(\lambda_{1},\mu_{1})\rangle_{\rho'}=\delta_{\rho\rho'}.
\end{equation}
When the Wigner-Eckart theorem~(\ref{9}) is applied, for such an operator, the sum over the outer multiplicity index reduces to a single term, and the SU(2)-RMEs of the Wigner operators are identified with RCCs:
\begin{multline}\label{eqn:wigner-su2-rme}
\rbrajh{(\lambda_{3},\mu_{3})}{\epsilon_{3}\Lambda_{3}}K_{\epsilon_{2}\Lambda_{2}}^{(\lambda_{2},\mu_{2})\rho}\rketjh{(\lambda_{1},\mu_{1})}{\epsilon_{1}\Lambda_{1}}\\
=\rwig{(\lambda_{1},\mu_{1})}{\epsilon_{1}\Lambda_{1}}{(\lambda_{2},\mu_{2})}{\epsilon_{2}\Lambda_{2}}{(\lambda_{3},\mu_{3})}{\epsilon_{3}\Lambda_{3}}_{\rho}.
\end{multline}

We can motivate how the BLH vanishing conditions~(\ref{blh}) might be enforced by relating the RCCs to RMEs of an operator, schematically, by considering an operator $K'^{(\lambda_{2},\mu_{2})\rho}$, defined by a ``stretched'' $\grpsu{3}$ coupling of $\eta$ copies of the $\grpsu{3}$ generator onto a Wigner operator:
\begin{equation}
  \label{eqn:k-schematic}
  K'^{(\lambda_{2},\mu_{2})\rho}
  =
  \bigl[K^{(\bar\lambda_{2},\bar\mu_{2})\rho}
\underbrace{\times C^{(1,1)}\cdots\times
      C^{(1,1)}}_{\text{$\eta$ times}} \bigr]^{(\lambda_{2},\mu_{2})},
\end{equation}
where, specifically, $K^{(\bar\lambda_{2},\bar\mu_{2})\rho}$ is a Wigner operator~(\ref{eqn:wigner-su3-rme}) for the coupling $(\lambda_{1},\mu_{1})\times(\bar\lambda_{2},\bar\mu_{2})\to(\lambda_{3},\mu_{3})$. Then, consider the $\grpsu{2}$-RME
\begin{equation}
  \label{eqn:rme-K-prime}
  \rbrajh{(\lambda_{3},\mu_{3})}{\epsilon_{3}\Lambda_{3}}K'^{(\lambda_{2},\mu_{2})\rho}_{\epsilon_{2}\Lambda_{2}}\rketjh{(\lambda_{1},\mu_{1})}{\epsilon_{1}\Lambda_{1}}.
\end{equation}
As an $\grpsu{3}$ tensor operator, the Wigner operator appearing on the right-hand side of~(\ref{eqn:k-schematic}), $K^{(\bar\lambda_{2},\bar\mu_{2})\rho}$, can change $\Lambda$ by at most the maximal $\Lambda$ appearing in the irrep $(\bar{\lambda}_2,\bar{\mu}_2)$, which is $\frac{1}{2}(\bar{\lambda}_2+\bar{\mu}_2)$.  Then, although the $\grpsu{3}$ generator $C^{(1,1)}_{\epsilon\Lambda}$ contains components with $\Lambda=0$, $1/2$, and $1$, the component with $\Lambda=1$ is simply the $\grpsu{2}$ generator, and therefore cannot change $\Lambda$ at all.  The components with $\Lambda=1/2$ can change $\Lambda$ by at most $\frac{1}{2}$.  Thus, $\eta$ successive applications give a total allowed change $|\Lambda_1-\Lambda_3|\leq \frac{1}{2}(\lambda_2+\mu_2-\eta_{\rm max}+\rho)$, and the RME~(\ref{eqn:rme-K-prime}) vanishes under exactly the same condition as the RCC with corresponding quantum numbers in~(\ref{blh}).

To derive a building-up recurrence relation\footnote{We provide here an alternate derivation of the building-up recurrence relation~(13) of Ref.~\cite{draayer}, avoiding any reference to projection quantum numbers, by use of identities for RMEs, RCCs, and recoupling coefficients.}  between the RCCs for successive couplings $(\lambda_{1},\mu_{1})\times(\lambda_{2}-1,\mu_{2}-1)\to(\lambda_{3},\mu_{3})$ and $(\lambda_{1},\mu_{1})\times(\lambda_{2},\mu_{2})\to(\lambda_{3},\mu_{3})$, we encode the selection rules induced by the action of the $\grpsu{3}$ generator, as in the schematic discussion above, by considering the RMEs of an operator $K'^{(\lambda_{2},\mu_{2})\rho}$, obtained by coupling a single factor of $C^{(1,1)}$ onto the Wigner operator $K^{(\lambda_{2}-1,\mu_{2}-1)\rho}$:
\begin{equation}\label{11}
  K'^{(\lambda_{2},\mu_{2})\rho}=\bigl[K^{(\lambda_{2}-1,\mu_{2}-1)\rho}\times C^{(1,1)}\bigr]^{(\lambda_{2},\mu_{2})},
\end{equation}
where the prime is used to denote the fact that $K'^{(\lambda_{2},\mu_{2})\rho}$ as thus defined is not itself, in general, a Wigner operator satisfying~(\ref{eqn:wigner-su3-rme}).\footnote{Operators or RCCs distinguished in Ref.~\cite{draayer} by a tilde on the outer multiplicity index are distinguished here by a prime on the operator or RCC itself, so that the symbol for the outer multiplicity index need not be construed to carry any meaning other than simply representing the integer value of the index.}

Since $K'^{(\lambda_{2},\mu_{2})\rho}$ is defined, in~(\ref{11}), as a coupled
product of two $\grpsu{3}$ tensor operators, its $\grpsu{3}$-RME may be
evaluated in terms of the RMEs of two operators separately, as well as an
$\grpsu{3}$ recoupling coefficient.  The appropriate
generalization~\cite{butler} of Racah's reduction
formula~\cite{racah1942:complex-spectra-part2-tensor-alg} from angular momentum
theory [see~(7.1.1) of Ref.~\cite{edmonds1960:am}], to $\grpsu{3}$ tensor
operators and RMEs, is given in~(B.23) of Ref.~\cite{mccoy2018:diss}.  The
$\grpsu{3}$-RME of $K^{(\lambda_{2}-1,\mu_{2}-1)\rho}$ is trivial, by
relation~(\ref{eqn:wigner-su3-rme}), and the $\grpsu{3}$-RME of the generator
reduces to~\cite{hecht}
\begin{multline}\label{eqn:generator-su3-rme}
  \langle(\lambda',\mu')||C^{(1,1)}||(\lambda,\mu)\rangle_{\rho}\\
  =
  \delta_{(\lambda',\mu'),(\lambda,\mu)}
  \delta_{\rho1}\langle(\lambda,\mu)||C^{(1,1)}||(\lambda,\mu)\rangle,
\end{multline}
so that the $\grpsu{3}$-RME of $K'^{(\lambda_{2},\mu_{2})\rho}$ is
\begin{multline}\label{intermezzo}
  \langle(\lambda_{3},\mu_{3})||K'^{(\lambda_{2},\mu_{2})\rho}||(\lambda_{1},\mu_{1})\rangle_{\rho'}\\
  =\langle(\lambda_{1},\mu_{1})||C^{(1,1)}||(\lambda_{1},\mu_{1})\rangle
  \\
  \times
  U[(\lambda_{1},\mu_{1})(1,1)(\lambda_{3},\mu_{3})\\
    (\lambda_2-1,\mu_2-1);(\lambda_{1},\mu_{1})1\rho(\lambda_2,\mu_2)_{-}\rho'].
\end{multline}

We seek, however, a relation among RCCs.  Evaluating the $\grpsu{2}$-RME, by the Wigner-Eckart theorem~(\ref{9}), involves multiplying the $\grpsu{3}$-RME on the left-hand side of~(\ref{intermezzo}) by an $\grpsu{3}$ RCC and summing over the multiplicity index:\footnote{It may be argued~\cite{draayer} that the summation over $\rho'$ in~(\ref{eqn:Kprime-we-cgp}) can be restricted to $\rho'\leq\rho$, by the null space properties of the Wigner operator, but that observation is not essential to the derivation of the recurrence relation below.}
\begin{multline}
  \label{eqn:Kprime-we-cgp}
  \rbrajh{(\lambda_{3},\mu_{3})}{\epsilon_{3}\Lambda_{3}}K'^{(\lambda_{2},\mu_{2})\rho}_{\epsilon_2\Lambda_2}\rketjh{(\lambda_1,\mu_1)}{\epsilon_1\Lambda_1}
  \\
  \begin{aligned}
    =&\sum_{\rho'}\langle(\lambda_{3},\mu_{3})||K'^{(\lambda_{2},\mu_{2})\rho}||(\lambda_{1},\mu_{1})\rangle_{\rho'}\\
    &\times\rwig{(\lambda_{1},\mu_{1})}{\epsilon_{1}\Lambda_{1}}{(\lambda_{2},\mu_{2})}{\epsilon_{2}\Lambda_{2}}{(\lambda_{3},\mu_{3})}{\epsilon_{3}\Lambda_{3}}_{\rho'}
    \\
    \equiv&\rwig{(\lambda_{1},\mu_{1})}{\epsilon_{1}\Lambda_{1}}{(\lambda_{2},\mu_{2})}{\epsilon_{2}\Lambda_{2}}{(\lambda_{3},\mu_{3})}{\epsilon_{3}\Lambda_{3}}'_{\rho}.
  \end{aligned}
\end{multline}
In the summation on the right-hand side of~(\ref{eqn:Kprime-we-cgp}) we take a linear combination of RCCs with coefficients depending only upon the outer multiplicity index. This linear combination may be taken as a (non-unitary) transformation~(\ref{eqn:cg-unitary-xform}) to a (non-orthonormal) set of RCCs for the coupling $(\lambda_{1},\mu_{1})\times(\lambda_{2},\mu_{2})\to(\lambda_{3},\mu_{3})$, which we denote by the primed RCCs in~(\ref{eqn:Kprime-we-cgp}).  However, as noted above, our immediate aim in the building-up process is merely to obtain a valid set of RCCs for one given value of the multiplicity index $\rho$, satisfying the BLH vanishing conditions~(\ref{blh}), without regard for orthogonality with respect to the RCCs obtained for other $\rho$, or for normalization.  Thus, it is sufficient if we derive a recurrence relation which yields these primed RCCs.

Similarly, multiplying the $\grpsu{3}$ recoupling coefficient on the right-hand side of~(\ref{intermezzo}) by an $\grpsu{3}$ RCC and summing over the multiplicity index, we recognize that we can apply the $\grpsu{3}$ recoupling coefficient identity~\cite{hecht,draayer}
\begin{multline}\label{22}
  \sum_{\rho_{1,23}}\rwig{(\lambda_{1},\mu_{1})}{\epsilon_{1}\Lambda_{1}}{(\lambda_{23},\mu_{23})}{\epsilon_{23}\Lambda_{23}}{(\lambda,\mu)}{\epsilon\Lambda}_{\rho_{1,23}}\\
  \times U[(\lambda_{1},\mu_{1})(\lambda_{2},\mu_{2})(\lambda,\mu)(\lambda_{3},\mu_{3});\\
    (\lambda_{12},\mu_{12})\rho_{12},\rho_{12,3}(\lambda_{23},\mu_{23})\rho_{23},\rho_{1,23}]\\
  =\sum_{\substack{\epsilon_{2}\Lambda_{2}\Lambda_{3}\Lambda_{12}\\(\epsilon_3\epsilon_{12})}}\rwig{(\lambda_{1},\mu_{1})}{\epsilon_{1}\Lambda_{1}}{(\lambda_{2},\mu_{2})}{\epsilon_{2}\Lambda_{2}}{(\lambda_{12},\mu_{12})}{\epsilon_{12}\Lambda_{12}}_{\rho_{12}}\\
  \times\rwig{(\lambda_{12},\mu_{12})}{\epsilon_{12}\Lambda_{12}}{(\lambda_{3},\mu_{3})}{\epsilon_{3}\Lambda_{3}}{(\lambda,\mu)}{\epsilon\Lambda}_{\rho_{12,3}}\\
  \times\rwig{(\lambda_{2},\mu_{2})}{\epsilon_{2}\Lambda_{2}}{(\lambda_{3},\mu_{3})}{\epsilon_{3}\Lambda_{3}}{(\lambda_{23},\mu_{23})}{\epsilon_{23}\Lambda_{23}}_{\rho_{23}}\\
  \times U(\Lambda_1\Lambda_2\Lambda\Lambda_3;\Lambda_{12}\Lambda_{23})
\end{multline}
to eliminate the $\grpsu{3}$ recoupling coefficient in favor of $\grpsu{3}$ RCCs and an $\grpsu{2}$ recoupling coefficient.  We thus obtain the recurrence relation~\cite{draayer}\footnote{Compared to~(13) of Ref.~\cite{draayer}, normalization factors arising from the $\grpsu{3}$-RMEs of $K$ and $K'$ are eliminated, by virtue of the choices of normalization in~(\ref{eqn:wigner-su3-rme}) and~(\ref{eqn:Kprime-we-cgp}).}
\begin{multline}\label{13}
  \rwig{(\lambda_{1},\mu_{1})}{\epsilon_{1}\Lambda_{1}}{(\lambda_{2},\mu_{2})}{\epsilon_{2}\Lambda_{2}}{(\lambda_{3},\mu_{3})}{\epsilon_{3}\Lambda_{3}}'_{\rho}\\
  =\langle(\lambda_{1},\mu_{1})||C^{(11)}||(\lambda_{1},\mu_{1})\rangle\\
  \times\sum_{\substack{\epsilon\Lambda\Lambda_{1}'\Lambda_{2}'\\(\epsilon_1'\epsilon_2')}}\rwig{(1,1)}{\epsilon\Lambda}{(\lambda_{2}-1,\mu_{2}-1)}{\epsilon_{2}'\Lambda_{2}'}{(\lambda_{2},\mu_{2})}{\epsilon_{2}\Lambda_{2}}\\
  \times\rwig{(\lambda_{1},\mu_{1})}{\epsilon_{1}\Lambda_{1}}{(1,1)}{\epsilon\Lambda}{(\lambda_{1},\mu_{1})}{\epsilon_{1}'\Lambda_{1}'}_{1}\\
  \times\rwig{(\lambda_{1},\mu_{1})}{\epsilon_{1}'\Lambda_{1}'}{(\lambda_{2}-1,\mu_{2}-1)}{\epsilon_{2}'\Lambda_{2}'}{(\lambda_{3},\mu_{3})}{\epsilon_{3}\Lambda_{3}}_{\rho}\\
  \times U(\Lambda_1\Lambda\Lambda_3\Lambda_2';\Lambda_1'\Lambda_2).
\end{multline}
The generator RME in~(\ref{13}) contributes only an overall normalization factor to the set of RCCs yielded by the recurrence relation, for the given $\rho$, and may thus be omitted in application of the recurrence relation, since normalization of this set will anyway later be enforced by Gram-Schmidt orthnormalization over the outer multiplicity index.

Note that the action of the generator in~(\ref{11}), which is ultimately responsible for imposing the BLH conditions~(\ref{blh}) on the difference in $\Lambda$, is encoded in the recurrence relation~(\ref{13}) through the appearance of a ``generator RCC''.  Since, equivalently to~(\ref{eqn:generator-su3-rme}) by applying the Wigner-Eckart theorem~(\ref{9}),
\begin{multline}\label{eqn:generator-su2-rme-same-irrep}
  \rbrajh{(\lambda_{1},\mu_{1})}{\epsilon_{1}'\Lambda_{1}'}C_{\epsilon\Lambda}^{(1,1)}\rketjh{(\lambda_{1},\mu_{1})}{\epsilon_{1}\Lambda_{1}}
  \\
  =\langle(\lambda_{1},\mu_{1})||C^{(1,1)}||(\lambda_{1},\mu_{1})\rangle\\
  \times\rwig{(\lambda_{1},\mu_{1})}{\epsilon_{1}\Lambda_{1}}{(1,1)}{\epsilon\Lambda}{(\lambda_{1},\mu_{1})}{\epsilon_{1}'\Lambda_{1}'}_{1},
\end{multline}
the appearance of this same RCC in~(\ref{13}) restricts $|\Lambda'_1-\Lambda_1|\leq 1/2$.  After a single application of this recurrence, $|\Lambda_1-\Lambda_3|$ may increase by at most $1/2$ relative to $|\Lambda_1'-\Lambda_3|$, and $\eta$ successive applications of the recurrence yields the BLH condition of~(\ref{blh}).

So that the recurrence~(\ref{13}) needs only to be applied to obtain a limited number of RCCs, and with the goal of evaluating extremal RCCs in mind, it is practical to restrict both $\epsilon_{2}\Lambda_{2}$ and $\epsilon_{3}\Lambda_{3}$ to be of the highest weight. This forces $\epsilon\Lambda$ and $\epsilon_{2}'\Lambda_{2}'$ to be of the highest weight as well. Then
\begin{equation}
  \rwig{(11)}{\epsilon^{\rm H}\Lambda^{\rm H}}{(\lambda_{2}-1,\mu_{2}-1)}{\epsilon_{2}'^{\rm H}\Lambda_{2}'^{\rm H}}{(\lambda_{2}\mu_{2})}{\epsilon_{2}^{\rm H}\Lambda_{2}^{\rm H}}=1,
\end{equation}
and one obtains the relation
\begin{multline}\label{17}
  \rwig{(\lambda_{1},\mu_{1})}{\epsilon_{1}\Lambda_{1}}{(\lambda_{2},\mu_{2})}{\epsilon_{2}^{\rm H}\Lambda_{2}^{\rm H}}{(\lambda_{3},\mu_{3})}{\epsilon_{3}^{\rm H}\Lambda_{3}^{\rm H}}'_{\rho}\\
  =\langle(\lambda_{1},\mu_{1})||C^{(11)}||(\lambda_{1},\mu_{1})\rangle\\
  \times\sum_{\Lambda_{1}'=\Lambda_1\pm\frac{1}{2}}\rwig{(\lambda_{1},\mu_{1})}{\epsilon_{1}\Lambda_{1}}{(1,1)}{-3,\frac{1}{2}}{(\lambda_{1},\mu_{1})}{\epsilon_{1}-3,\Lambda_{1}'}_{1}\\
  \times\rwig{(\lambda_{1},\mu_{1})}{\epsilon_{1}-3,\Lambda_{1}'}{(\lambda_{2}-1,\mu_{2}-1)}{\epsilon_{2}'^{\rm H}\Lambda_{2}'^{\rm H}}{(\lambda_{3},\mu_{3})}{\epsilon_{3}^{\rm H}\Lambda_{3}^{\rm H}}_{\rho}\\
  \times U\bigl(\Lambda_1\tfrac{1}{2}\Lambda_3^{\rm H}\Lambda_2'^{\rm H};\Lambda_1'\Lambda_2^{\rm H}\bigr),
\end{multline}
where $\epsilon_{2}'^{\rm H}\Lambda_{2}'^{\rm H}$ are the highest-weight quantum numbers for the irrep $(\lambda_{2}-1,\mu_{2}-1)$.  An analytic expression for the generator RCC, that is, involving the $(1,1)$ irrep, in~(\ref{17}), is available in Ref.~\cite{hecht}, and analytic expressions are available for the $\grpsu{2}$ recoupling coefficients as well~\cite{var}.  The generator RME in~(\ref{17}) again plays the role of a normalization factor, which may be omitted in anticipation of subsequent Gram-Schmidt orthonormalization.

\subsubsection{Draayer-Akiyama algorithm}
\label{sec:algorithm:canonical:da}

The DA algorithm~\cite{draayer} for calculation of the $\grpsu{3}\supset\grpu{1}\times\grpsu{2}$ RCCs then proceeds as follows.  First, a set of extremal RCCs (\textit{i.e.}, having extremal $\epsilon_3\Lambda_3$) is obtained, without concern for orthonormality with respect to $\rho$.  Independently, for each $\rho=1,\ldots,\rho_{\rm max}$:

\paragraph{Step 1.} 
The coefficients
\begin{equation}\label{step1}
\rwig{(\lambda_{1},\mu_{1})}{\epsilon_{1}^{\rm H}\Lambda_{1}^{\rm H}}{(\bar{\lambda}_{2},\bar{\mu}_{2})}{\bar{\epsilon}_{2}\bar{\Lambda}_{2}}{(\lambda_{3},\mu_{3})}{\epsilon_{3}^{\rm H}\Lambda_{3}^{\rm H}}_{\rho},
\end{equation}
are generated, from the explicit expression~(20) in Ref.~\cite{draayer}.\footnote{See also~(12) of Ref.~\cite{asherova1997:u3-outer-multiplicity} for an alternative expression for the RCCs~(\ref{step1}).  While we retain this first step from Ref.~\cite{draayer} for completeness, note that the final results for the extremal RCCs obtained below, after orthonormalization, are independent of the values provided for the RCCs~(\ref{step1}) in Step~1.  These serve as seeds for the recurrence relation stemming from the method of infinitesimal generators in Step~2, which guarantees a valid set of RCCs, and then for the building-up recurrence~(\ref{17}) in Step~3, which enforces the BLH resolution of the outer multiplicity.  Together with the imposed orthonormalization and phase convention, these conditions uniquely determine the RCCs.  It is only necessary that the seed values provided in Step~1 provide a linearly independent (and thus complete) set of RCCs entering into the orthonormalization process.}

\paragraph{Step 2.} 
From the coefficients~(\ref{step1}), the coefficients
\begin{equation}\label{step2}
\rwig{(\lambda_{1},\mu_{1})}{\epsilon_{1}\Lambda_{1}}{(\bar{\lambda}_{2},\bar{\mu}_{2})}{\bar{\epsilon}_{2}^{\rm H}\bar{\Lambda}_{2}^{\rm H}}{(\lambda_{3},\mu_{3})}{\epsilon_{3}^{\rm H}\Lambda_{3}^{\rm H}}_{\rho},
\end{equation}
with $\bar{\epsilon}_{2}\bar{\Lambda}_{2}$ of highest weight, are generated using the recurrence relation~(\ref{19}) from the method of infinitesimal generators, which reduces for this purpose to the form~(\ref{21}).

\paragraph{Step 3.} 
From the coefficients~(\ref{step2}), the coefficients
\begin{equation}\label{step3}
\rwig{(\lambda_{1},\mu_{1})}{\epsilon_{1}\Lambda_{1}}{(\lambda_{2},\mu_{2})}{\epsilon_{2}^{\rm H}\Lambda_{2}^{\rm H}}{(\lambda_{3},\mu_{3})}{\epsilon_{3}^{\rm H}\Lambda_{3}^{\rm H}}'_{\rho},
\end{equation}
with $\epsilon_{2}\Lambda_{2}$ of highest weight, are generated using the building-up recurrence relation~(\ref{17}).

\paragraph{Step 4.} 
From the coefficients~(\ref{step3}), the remaining extremal coefficients (\textit{i.e.}, with $\epsilon_{2}\Lambda_{2}$ not of highest weight)
\begin{equation}\label{step4}
\rwig{(\lambda_{1},\mu_{1})}{\epsilon_{1}\Lambda_{1}}{(\lambda_{2},\mu_{2})}{\epsilon_{2}\Lambda_{2}}{(\lambda_{3},\mu_{3})}{\epsilon_{3}^{\rm H}\Lambda_{3}^{\rm H}}'_{\rho}
\end{equation}
are generated by again using the recurrence relation~(\ref{19}) from the method of infinitesimal generators, which reduces for this purpose to the form~(\ref{18}).

The sets of extremal RCCs obtained in this way for different $\rho$ are then orthonormalized with respect to $\rho$ using the Gram-Schmidt procedure.  As noted in Sec.~\ref{sec:algorithm:canonical:building-up}, this orthonormalization must be carried out in order of increasing $\rho$ to preserve the BLH constraints.  The phase convention of Ref.~\cite{hecht} is imposed on the resulting orthonormal RCCs:\footnote{In Ref.~\cite{hecht}, the condition~(\ref{phase}) is formulated as $\rwig{(\lambda_{1},\mu_{1})}{\epsilon_{1}^{\rm L}\Lambda_{1}^{\rm L}}{(\lambda_{2},\mu_{2})}{\epsilon_{2}\Lambda_{2,\rm max}}{(\lambda_{3},\mu_{3})}{\epsilon_{3}^{\rm L}\Lambda_{3}^{\rm L}}_{\rho}>0$, from which~(\ref{phase}) can be obtained using the symmetry property~(\ref{conjug}).}
\begin{multline}\label{phase}
\rwig{(\lambda_{1},\mu_{1})}{\epsilon_{1}^{\rm H}\Lambda_{1}^{\rm H}}{(\lambda_{2},\mu_{2})}{\epsilon_{2}\Lambda_{2,\rm max}}{(\lambda_{3},\mu_{3})}{\epsilon_{3}^{\rm H}\Lambda_{3}^{\rm H}}_{\rho}\\
\times(-1)^{\varphi+\rho_{\rm max}-\rho+\frac{\lambda_{1}}{2}+\Lambda_{2,\rm max}-\frac{\lambda_{3}}{2}}>0,
\end{multline}
where $\varphi=\lambda_{1}+\lambda_{2}-\lambda_{3}+\mu_{1}+\mu_{2}-\mu_{3}$.

The RCCs with non-extremal $\epsilon_{3}\Lambda_{3}$ are obtained from those with $\epsilon_{3}^{\rm H}\Lambda_{3}^{\rm H}$ recursively, again using the recurrence relation~(\ref{19}) from the method of infinitesimal generators. However, this is not always the shortest path. In present implementation, if $2\epsilon_{3}>\lambda_{3}-\mu_{3}$ we instead recurse from the RCCs with $\epsilon_{3}^{\rm L}\Lambda_{3}^{\rm L}$ via~(\ref{19conjug}).

The RCCs with $\epsilon_{3}^{\rm L}\Lambda_{3}^{\rm L}$ are related to those with $\epsilon_{3}^{\rm H}\Lambda_{3}^{\rm H}$ by symmetry property~\cite{draayer}
\begin{multline}\label{conjug}
\rwig{(\lambda_{1},\mu_{1})}{\epsilon_{1}\Lambda_{1}}{(\lambda_{2},\mu_{2})}{\epsilon_{2}\Lambda_{2}}{(\lambda_{3},\mu_{3})}{\epsilon_{3}^{\rm L}\Lambda_{3}^{\rm L}}_{\rho}\\
=(-1)^{\varphi+\rho_{\rm max}-\rho+\Lambda_{1}+\Lambda_{2}-\frac{\mu_{3}}{2}}\\
\times\rwig{(\mu_{1},\lambda_{1})}{-\epsilon_{1}\Lambda_{1}}{(\mu_{2},\lambda_{2})}{-\epsilon_{2}\Lambda_{2}}{(\mu_{3},\lambda_{3})}{\epsilon_{3}^{\rm H}\Lambda_{3}^{\rm H}}_{\rho}.
\end{multline}
Thus, we calculate the RCCs with $\epsilon_{3}^{\rm L}\Lambda_{3}^{\rm L}$ by first calculating the RCCs with $\lambda_{i}$ and $\mu_{i}$ swapped and with $\epsilon_{3}^{\rm H}\Lambda_{3}^{\rm H}$, and then applying the symmetry transformation~(\ref{conjug}).

\subsection{$\grpsu{3}\supset\grpso{3}$ reduced coupling coefficients}
\label{sec:algorithm:physical}

The $\grpsu{3}\supset\grpso{3}$ coupling coefficients can be obtained from the $\grpsu{3}\supset\grpu{1}\times\grpsu{2}$ coupling coefficients by a straightforward basis transformation in the irreps:
\begin{multline}\label{eqn:su3-so3-cq-xform}
\wig{(\lambda_{1},\mu_{1})}{\kappa_{1}L_{1}M_{1}}{(\lambda_{2},\mu_{2})}{\kappa_{2}L_{2}M_{2}}{(\lambda_{3},\mu_{3})}{\kappa_{3}L_{3}M_{3}}_{\rho}\\
=\sum_{\epsilon_{1}\Lambda_{1}M_{\Lambda_{1}}}\sum_{\epsilon_{2}\Lambda_{2}M_{\Lambda_{2}}}\sum_{\epsilon_{3}\Lambda_{3}M_{\Lambda_{3}}}\overlapjh{(\lambda_{1},\mu_{1})}{\epsilon_{1}\Lambda_{1}M_{\Lambda_{1}}}{(\lambda_{1},\mu_{1})}{\kappa_{1}L_{1}M_{1}}\\
\times\overlapjh{(\lambda_{2},\mu_{2})}{\epsilon_{2}\Lambda_{2}M_{\Lambda_{2}}}{(\lambda_{2},\mu_{2})}{\kappa_{2}L_{2}M_{2}}\overlapjh{(\lambda_{3},\mu_{3})}{\epsilon_{3}\Lambda_{3}M_{\Lambda_{3}}}{(\lambda_{3},\mu_{3})}{\kappa_{3}L_{3}M_{3}}\\
\times\wig{(\lambda_{1},\mu_{1})}{\epsilon_{1}\Lambda_{1}M_{\Lambda_{1}}}{(\lambda_{2},\mu_{2})}{\epsilon_{2}\Lambda_{2}M_{\Lambda_{2}}}{(\lambda_{3},\mu_{3})}{\epsilon_{3}\Lambda_{3}M_{\Lambda_{3}}}_{\rho},
\end{multline}
where the transformation brackets are given by~(\ref{ortoc}).  However, the summation in~(\ref{eqn:su3-so3-cq-xform}) involves the full set of $\grpsu{3}\supset\grpu{1}\times\grpsu{2}$ RCCs arising in the coupling $(\lambda_1,\mu_1)\times(\lambda_2,\mu_2)\to(\lambda_3,\mu_3)$.

A formula that is more practical for computational purposes can instead be obtained by first evaluating a set of non-orthogonal RCCs, yielding Elliott basis states for the product irrep.
From the definition~(\ref{-proj}) of an Elliott basis state, these are defined by taking the inner product
\begin{multline}
\wig{(\lambda_{1},\mu_{1})}{\kappa_{1}L_{1}M_{1}}{(\lambda_{2},\mu_{2})}{\kappa_{2}L_{2}M_{2}}{(\lambda_{3},\mu_{3})}{K_{3}L_{3}M_{3}}_{\rho}\\
=\left\langle\begin{array}{cc}
(\lambda_{1},\mu_{1}) & (\lambda_{2},\mu_{2}) \\
\kappa_{1}L_{1}M_{1} & \kappa_{2}L_{2}M_{2} \\
\end{array}\right|P_{M_{3}K_{3}}^{L_{3}}\left|\begin{array}{c}
(\lambda_{3},\mu_{3}) \\
\epsilon_{3}^{\rm E}\Lambda_{3}^{\rm E}M_{\Lambda_{3}}^{\rm E} \\
\end{array}\right\rangle_{\rho}.
\end{multline}
Acting with the projection operator to the left~\cite{draayer,draayer2} yields the result
\begin{multline}\label{31}
\wig{(\lambda_{1},\mu_{1})}{\kappa_{1}L_{1}M_{1}}{(\lambda_{2},\mu_{2})}{\kappa_{2}L_{2}M_{2}}{(\lambda_{3},\mu_{3})}{K_{3}L_{3}M_{3}}_{\rho}\\
=\wig{L_{1}}{M_{1}}{L_{2}}{M_{2}}{L_{3}}{M_{3}}\sum_{\substack{\Lambda_{1}M_{\Lambda_{1}}M_{1}'\epsilon_{2}\Lambda_{2}\\(\epsilon_1M_{\Lambda_2}M_2')}}\wig{L_{1}}{M_{1}'}{L_{2}}{M_{2}'}{L_{3}}{K_{3}}\\
\times\overlapjh{(\lambda_{1},\mu_{1})}{\epsilon_{1}\Lambda_{1}M_{\Lambda_{1}}}{(\lambda_{1},\mu_{1})}{\kappa_{1}L_{1}M_{1}'}\overlapjh{(\lambda_{2},\mu_{2})}{\epsilon_{2}\Lambda_{2}M_{\Lambda_{2}}}{(\lambda_{2},\mu_{2})}{\kappa_{2}L_{2}M_{2}'}\\
\times\wig{(\lambda_{1},\mu_{1})}{\epsilon_{1}\Lambda_{1}M_{\Lambda_{1}}}{(\lambda_{2},\mu_{2})}{\epsilon_{2}\Lambda_{2}M_{\Lambda_{2}}}{(\lambda_{3},\mu_{3})}{\epsilon_{3}^{\rm E}\Lambda_{3}^{\rm E}M_{\Lambda_{3}}^{\rm E}}_{\rho},
\end{multline}
where the choice of extremal state is given by~(\ref{e}). Note that the summation in~(\ref{31}) now involves only $\grpsu{3}\supset\grpu{1}\times\grpsu{2}$ coupling coefficients with extremal $\epsilon_{3}\Lambda_{3}M_{\Lambda_{3}}$.  To recast~(\ref{31}) as a relation among RCCs, the $\grpsu{3}$ coupling coefficients are factored via~(\ref{racfac}) and~(\ref{eqn:factorization:physical}), yielding a formula~\cite{draayer} for non-orthonormal $\grpsu{3}\supset\grpso{3}$ RCCs in terms of extremal $\grpsu{3}\supset\grpu{1}\times\grpsu{2}$ RCCs:
\begin{multline}\label{31m}
\rwig{(\lambda_{1},\mu_{1})}{\kappa_{1}L_{1}}{(\lambda_{2},\mu_{2})}{\kappa_{2}L_{2}}{(\lambda_{3},\mu_{3})}{K_{3}L_{3}}_{\rho}\\
=\sum_{\substack{\Lambda_{1}M_{\Lambda_{1}}M_{1}'\epsilon_{2}\Lambda_{2}\\(\epsilon_1M_{\Lambda_2}M_2')}}\wig{L_{1}}{M_{1}'}{L_{2}}{M_{2}'}{L_{3}}{K_{3}}\wig{\Lambda_{1}}{M_{\Lambda_{1}}}{\Lambda_{2}}{M_{\Lambda_{2}}}{\Lambda_{3}^{\rm E}}{M_{\Lambda_{3}}^{\rm E}}\\
\times\overlapjh{(\lambda_{1},\mu_{1})}{\epsilon_{1}\Lambda_{1}M_{\Lambda_{1}}}{(\lambda_{1},\mu_{1})}{\kappa_{1}L_{1}M_{1}'}\overlapjh{(\lambda_{2},\mu_{2})}{\epsilon_{2}\Lambda_{2}M_{\Lambda_{2}}}{(\lambda_{2},\mu_{2})}{\kappa_{2}L_{2}M_{2}'}\\
\times\rwig{(\lambda_{1},\mu_{1})}{\epsilon_{1}\Lambda_{1}}{(\lambda_{2},\mu_{2})}{\epsilon_{2}\Lambda_{2}}{(\lambda_{3},\mu_{3})}{\epsilon_{3}^{\rm E}\Lambda_{3}^{\rm E}}_{\rho}.
\end{multline}
Again, the transformation brackets between the orthonormal basis states reducing the canonical and angular momentum group chains are obtained using~(\ref{ortoc}).

Once the non-orthonormal $\grpsu{3}\supset\grpso{3}$ RCCs are obtained, using~(\ref{31m}), subsequent orthonormalization in the representation space of ($\lambda_{3},\mu_{3}$) with respect to the inner multiplicity label yields orthonormal $\grpsu{3}\supset\grpso{3}$ RCCs:
\begin{multline}\label{orto}
\rwig{(\lambda_{1},\mu_{1})}{\kappa_{1}L_{1}}{(\lambda_{2},\mu_{2})}{\kappa_{2}L_{2}}{(\lambda_{3},\mu_{3})}{\kappa_{3}L_{3}}_{\rho}\\
=\sum_{j=1}^{\kappa_{3}}O^{(\lambda_{3},\mu_{3})L_{3}}_{\kappa_{3}j}\rwig{(\lambda_{1},\mu_{1})}{\kappa_{1}L_{1}}{(\lambda_{2},\mu_{2})}{\kappa_{2}L_{2}}{(\lambda_{3},\mu_{3})}{K_{3,j}L_{3}}_{\rho},
\end{multline}
where the orthonormalization matrix $O^{(\lambda_3,\mu_3)L_3}$ is given by~(\ref{6a})--(\ref{6c}).

\subsection{$\grpsu{3}$ recoupling coefficients}
\label{sec:algorithm:recoupling}

Once we have the $\grpsu{3}$ RCCs (for the canonical group chain), we can calculate the $U$ and $Z$ recoupling coefficients by solving systems of linear equations involving these RCCs. These equations can be obtained by a generalization of the corresponding equations from the case of $\grpsu{2}$ coefficients~\cite{hecht}.

For the $U$ recoupling coefficients
\begin{multline*}
    U[(\lambda_{1},\mu_{1})(\lambda_{2},\mu_{2})(\lambda,\mu)(\lambda_{3},\mu_{3});
      \\
      (\lambda_{12},\mu_{12})\rho_{12},\rho_{12,3}(\lambda_{23},\mu_{23})\rho_{23},\rho_{1,23}],
\end{multline*}
these equations are given by~(\ref{22}).  A separate set of equations must be solved for each set of values of $\rho_{12}$, $\rho_{12,3}$, and $\rho_{23}$.  Taking $\rho_{1,23,\rm max}$ different values of $\Lambda_{23}$, while holding $\epsilon_1\Lambda_1$ and $\epsilon \Lambda$ (and thus $\epsilon_{23}$) fixed, one obtains from~(\ref{22}) a system of $\rho_{1,23,\rm max}$ linear equations for $\rho_{1,23,\rm max}$ different $U$ coefficients.

Recall that, in the DA algorithm, RCCs with non-extremal $\epsilon\Lambda$ in the coupled irrep must be calculated from extremal RCCs by recurrence (Sect.~\ref{sec:algorithm:canonical:da}).  To avoid unnecessary calculations of non-extremal RCCs, it is practical to choose $\epsilon_{1}\Lambda_{1}$ and $\epsilon\Lambda$ in~(\ref{22}) to be of the highest weight.  This choice, along with the symmetry property~(\ref{sympro}), yields the system of linear equations~(\ref{22m}) for the $U$ coefficients.  Note that, in~(\ref{22m}), three of the RCCs are extremal, and only one remaining RCC is non-extremal.

The $Z$ coefficients
\begin{multline*}
    Z[(\lambda_{2},\mu_{2})(\lambda_{1},\mu_{1})(\lambda,\mu)(\lambda_{3},\mu_{3});\\
      (\lambda_{12},\mu_{12})\rho_{12},\rho_{12,3}(\lambda_{13},\mu_{13})\rho_{13},\rho_{13,2}]
\end{multline*}
can be obtained similarly, by solving the system of linear equations~\cite{millener}
\begin{multline}\label{mil}
\sum_{\rho_{13,2}}\rwig{(\lambda_{13},\mu_{13})}{\epsilon_{13}\Lambda_{13}}{(\lambda_{2},\mu_{2})}{\epsilon_{2}\Lambda_{2}}{(\lambda,\mu)}{\epsilon\Lambda}_{\rho_{13,2}}\\
\times Z[(\lambda_{2},\mu_{2})(\lambda_{1},\mu_{1})(\lambda,\mu)(\lambda_{3},\mu_{3});\\
(\lambda_{12},\mu_{12})\rho_{12},\rho_{12,3}(\lambda_{13},\mu_{13})\rho_{13},\rho_{13,2}]\\
=\sum_{\substack{\epsilon_{1}\Lambda_{1}\Lambda_{3}\Lambda_{12}\\(\epsilon_3\epsilon_{12})}}\rwig{(\lambda_{1},\mu_{1})}{\epsilon_{1}\Lambda_{1}}{(\lambda_{3},\mu_{3})}{\epsilon_{3}\Lambda_{3}}{(\lambda_{13},\mu_{13})}{\epsilon_{13}\Lambda_{13}}_{\rho_{13}}\\
\times\rwig{(\lambda_{1},\mu_{1})}{\epsilon_{1}\Lambda_{1}}{(\lambda_{2},\mu_{2})}{\epsilon_{2}\Lambda_{2}}{(\lambda_{12},\mu_{12})}{\epsilon_{12}\Lambda_{12}}_{\rho_{12}}\\
\times\rwig{(\lambda_{12},\mu_{12})}{\epsilon_{12}\Lambda_{12}}{(\lambda_{3},\mu_{3})}{\epsilon_{3}\Lambda_{3}}{(\lambda,\mu)}{\epsilon\Lambda}_{\rho_{12,3}}\\
\times(-1)^{\Lambda_{1}+\Lambda-\Lambda_{12}-\Lambda_{13}}U(\Lambda_2\Lambda_1\Lambda\Lambda_3;\Lambda_{12}\Lambda_{13}).
\end{multline}
A separate set of equations must be solved for each set of values of $\rho_{12}$, $\rho_{12,3}$, and $\rho_{13}$.  Taking $\rho_{13,2,\rm max}$ different values of $\Lambda_{2}$, while holding $\epsilon_{13}\Lambda_{13}$ and $\epsilon \Lambda$ (and thus $\epsilon_{2}$) fixed, one obtains from~(\ref{mil}) a system of $\rho_{13,2,\rm max}$ linear equations for $\rho_{13,2,\rm max}$ different $Z$ coefficients.  It is similarly practical to choose $\epsilon_{13}\Lambda_{13}$ and $\epsilon\Lambda$ to be of the highest weight in~(\ref{mil}), yielding the system of equations~(\ref{z}) for the $Z$ coefficients.

A 9-($\lambda,\mu$) coefficient is calculated as a sum of products of one $Z$ and two $U$ coefficients~\cite{millener}:
\begin{multline}\label{9lm}
\left[\begin{array}{cccc}
(\lambda_{1},\mu_{1}) & (\lambda_{2},\mu_{2}) & (\lambda_{12},\mu_{12}) & \rho_{12} \\
(\lambda_{3},\mu_{3}) & (\lambda_{4},\mu_{4}) & (\lambda_{34},\mu_{34}) & \rho_{34} \\
(\lambda_{13},\mu_{13}) & (\lambda_{24},\mu_{24}) & (\lambda,\mu) & \rho_{13,24} \\
\rho_{13} & \rho_{24} & \rho_{12,34} &  \\
\end{array}\right]\\
=\sum_{\substack{\lambda_{0}\mu_{0}\rho_{13,2}\\\rho_{04}\rho_{12,3}}}U[(\lambda_{13},\mu_{13})(\lambda_{2},\mu_{2})(\lambda,\mu)(\lambda_{4},\mu_{4});\\
(\lambda_{0},\mu_{0})\rho_{13,2},\rho_{04}(\lambda_{24},\mu_{24})\rho_{24}\rho_{13,24}]\\
\times Z[(\lambda_{2},\mu_{2})(\lambda_{1},\mu_{1})(\lambda_{0},\mu_{0})(\lambda_{3},\mu_{3});\\
(\lambda_{12},\mu_{12})\rho_{12},\rho_{12,3}(\lambda_{13},\mu_{13})\rho_{13}\rho_{13,2}]\\
\times U[(\lambda_{12},\mu_{12})(\lambda_{3},\mu_{3})(\lambda,\mu)(\lambda_{4},\mu_{4});\\
(\lambda_{0},\mu_{0})\rho_{12,3},\rho_{04}(\lambda_{34},\mu_{34})\rho_{34}\rho_{12,34}].
\end{multline}
 \section{Structure, implementation details, and usage of the library}
\label{sec:implementation}

In this section we provide an overview of the \texttt{ndsu3lib}, including the code organization, details of implementation, and external library dependencies.  The code is orgnaized into four modules: (1) \texttt{ndsu3lib\_tools} which contains subroutines for initialization and finalization of the library and for evaluating outer and inner multiplicities, (2) \texttt{ndsu3lib\_coupling\_canonical} which contains subroutines for calculation of $\grpsu{3}\supset\grpu{1}\times\grpsu{2}$ RCCs, (3) \texttt{ndsu3lib\_coupling\_su3so3} which contains subroutines for calculation of $\grpsu{3}\supset\grpso{3}$ RCCs, and (4) \texttt{ndsu3lib\_recoupling} which contains subroutines for calculation of $\grpsu{3}$ recoupling coefficients.

\begin{figure*}
\includegraphics[width=\linewidth]{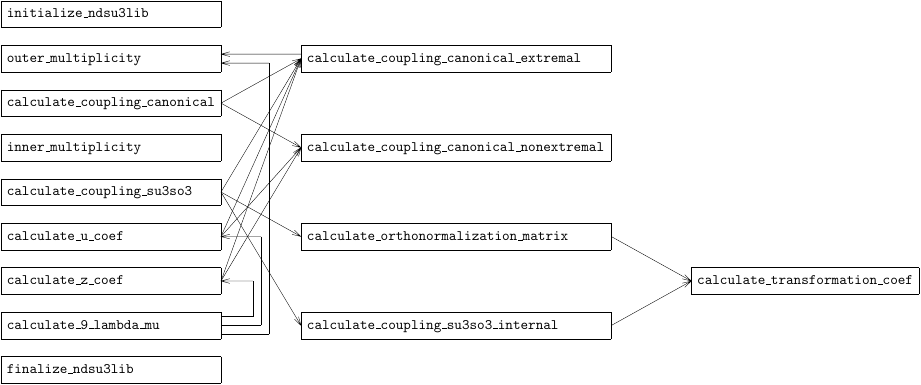}
\caption{Calling sequence of the subroutines and functions in
  \texttt{ndsu3lib}. Arrows point from subroutines to the subroutines or functions
  they call. The subroutines and functions at far left are to be called by the
  user.}
\label{fig:i}
\end{figure*}

The specific subroutines and functions in each module are given in Tables~\ref{tab:list0}--\ref{tab:list3}, and their calling sequence is depicted in Fig.~\ref{fig:i}. We distinguish between subroutines called by the user and internal subroutines that are not a part of the user interface. More details about usage and implemenation of the subrotuines and functions are given in the corresponding subsections below.

Examples of usage of the library are provided in the program \texttt{ndsu3lib\_example}. The program tabulates RCCs and recoupling coefficients for a choice of quantum numbers. The successful output of the program can be found in the file \texttt{example\_output.txt}.

A C/C++ header file \texttt{ndsu3lib.h} is provided to facilitate calling \texttt{ndsu3lib} from C or C++ code.  This header file provides wrappers to the subroutines and functions that form the \texttt{ndsu3lib} user interface. A C port \texttt{ndsu3lib\_example\_c} and C++ port \texttt{ndsu3lib\_example\_cpp} of the aforementioned example program are provided, demonstrating usage of the wrappers.

Configuration files are provided for compiling the library and associated example programs with the CMake build system.  Compilation instructions may be found in the file \texttt{INSTALL.md}.

The \texttt{ndsu3lib} library requires an external library for calculation of $\grpsu{2}$ coupling coefficients and 6$j$ symbols, and can be configured to use either the GNU Scientific Library (GSL) or the WIGXJPF library~\cite{wigxjpf} for this purpose. The choice between these two libraries is made at compilation. To avoid loss of precision when calculating $\grpsu{3}\supset\grpso{3}$ RCCs, \texttt{ndsu3lib} may also be configured to use multiprecision arithmetic, in which case the external library MPFUN2020~\cite{mpfun} is also required.

Since \texttt{ndsu3lib} implements the same underlying DA algorithm (section~\ref{sec:algorithm:canonical:da}) as the AD and \texttt{SU3lib} libraries, all three libraries yield consistent sets of coupling (and recoupling) coefficients, differing only in matters of numerical precision (section~\ref{sec:precision}).  Namely, all three libraries follow the same phase convention~(\ref{phase}) of Hecht~\cite{hecht} and the same BLH prescription~(\ref{blh}) for the resolution of outer multiplicities in generating the canonical RCCs, from which the recoupling coefficients follow.  Furthermore, in generating the $\grpsu{3}\supset\grpso{3}$ reduced coupling coefficients, all three libraries follow the same prescription for resolution of the inner multiplicities (section~\ref{sec:algorithm:physical}), based on Gram-Schmidt orthonormalization~(\ref{ortos}) of the Elliott basis.

\begin{table*}
\caption{Subroutines and functions in the module \texttt{ndsu3lib\_tools}.}
\label{tab:list0}
\begin{tabular*}{\textwidth}{@{\extracolsep{\fill}}lll@{}}
\hline Subroutine or function & Task & Implemented formulae \\
\hline
\texttt{initialize\_ndsu3lib} & Allocate and precalculate arrays and initialize
WIGXJPF & (\ref{i1}),~(\ref{s1}),~(\ref{s2}),~(\ref{i2}),~(\ref{i3}) \\
 \texttt{finalize\_ndsu3lib} & Deallocate memory &
\\
 \texttt{outer\_multiplicity} & Calculate multiplicity of $\grpsu{3}$ coupling
& Proposition 7(a) in~\cite{oreilly} \\
 \texttt{inner\_multiplicity} & Calculate
multiplicity of $L$ within $(\lambda,\mu)$ & (\ref{kappamax}) \\
\hline
\end{tabular*}
\end{table*}

\begin{table*}
\caption{Subroutines in the module \texttt{ndsu3lib\_coupling\_canonical}. The internal
  subroutines that are not a part of the user interface are denoted by asterisks.}
\label{tab:list}
\begin{tabular*}{\textwidth}{@{\extracolsep{\fill}}lll@{}}
\hline Subroutine & Task & Implemented formulae \\
\hline
\texttt{calculate\_coupling\_canonical\_extremal}* & Calculate extremal
$\grpsu{3}\supset\grpu{1}\times\grpsu{2}$ RCCs &
(\ref{17}),~(\ref{conjug}),~(\ref{21}),~(\ref{18}) \\
 \texttt{calculate\_coupling\_canonical\_nonextremal}* & Calculate non-extremal
$\grpsu{3}\supset\grpu{1}\times\grpsu{2}$ & (\ref{19}),~(\ref{19conjug})
\\
 & RCCs &
\\
 \texttt{calculate\_coupling\_canonical} & Calculate
$\grpsu{3}\supset\grpu{1}\times\grpsu{2}$ RCCs & \\
\hline
\end{tabular*}
\end{table*}

\begin{table*}
\caption{Subroutines in the module \texttt{ndsu3lib\_coupling\_su3so3}. The internal
  subroutines that are not a part of the user interface are denoted by asterisks.}
\label{tab:list2}
\begin{tabular*}{\textwidth}{@{\extracolsep{\fill}}lll@{}}
\hline Subroutine & Task & Implemented formulae \\
\hline
\texttt{calculate\_transformation\_coef}* & Calculate inner product of
$\grpsu{3}\supset\grpu{1}\times\grpsu{2}$ and & (\ref{26}) \\
 & Elliott basis
states & \\
 \texttt{calculate\_orthonormalization\_matrix}* & Calculate
orthonormalization matrix $O^{(\lambda,\mu)L}$ &
(\ref{6a}),~(\ref{6b}),~(\ref{6c}),~(\ref{6d})
\\
 \texttt{calculate\_coupling\_su3so3\_internal}* & Internal subroutine for
calculation of $\grpsu{3}\supset\grpso{3}$ &
(\ref{ortoc}),~(\ref{31m}),~(\ref{orto}) \\
 & RCCs &
\\
 \texttt{calculate\_coupling\_su3so3} & Calculate $\grpsu{3}\supset\grpso{3}$
RCCs & \\
\hline
\end{tabular*}
\end{table*}

\begin{table*}
\caption{Subroutines in the module \texttt{ndsu3lib\_recoupling}.}
\label{tab:list3}
\begin{tabular*}{\textwidth}{@{\extracolsep{\fill}}lll@{}}
\hline Subroutine & Task & Implemented formulae \\
\hline
\texttt{calculate\_u\_coef} & Calculate $U$ recoupling coefficients &
(\ref{22m}) \\
 \texttt{calculate\_z\_coef} & Calculate $Z$ recoupling
coefficients & (\ref{z}) \\
 \texttt{calculate\_9\_lambda\_mu} & Calculate
9-($\lambda,\mu$) coefficients & (\ref{9lm}) \\
\hline
\end{tabular*}
\end{table*}

\subsection{Module \texttt{ndsu3lib\_tools}}
\label{sec:implementation:tools}

Before a program first invokes \texttt{ndsu3lib} to calculate $\grpsu{3}$ coupling or recoupling coefficients, it must first initialize the library, by calling the subroutine \texttt{initialize\_ndsu3lib}. To increase speed and avoid loss of precision, this subroutine allocates and recursively precalculates arrays containing binomial coefficients and, optionally, factors $I(i,j,k)$ and $S(i,j,k)$. These factors are defined in~(\ref{i}) and~(\ref{s}) and are needed only if $\grpsu{3}\supset\grpso{3}$ RCCs are to be calculated. If the WIGXJPF library is being used, then \texttt{initialize\_ndsu3lib} also initializes WIGXJPF. While \texttt{ndsu3lib} is safe to use in OpenMP multithreaded applications, the subroutine \texttt{initialize\_ndsu3lib} should be called separately by each thread. In such applications the arrays of binomial coefficients and factors $I(i,j,k)$ and $S(i,j,k)$ are shared, and only one thread initializes them.

If the calling program has no further need for the \texttt{ndsu3lib} library, it may call the subroutine \texttt{finalize\_ndsu3lib}, to release the memory used for precomputed coefficients. In OpenMP multithreaded applications, this subroutine should be called by each thread.

The function \texttt{outer\_multiplicity} calculates the multiplicity of a given $\grpsu{3}$ coupling, implementing the algorithm of O'Reilly~\cite{oreilly}.

The function \texttt{inner\_multiplicity} calculates the number of occurences of a given $L$ within a given $\grpsu{3}$ irrep.

\subsection{Module \texttt{ndsu3lib\_coupling\_canonical}}

The user calculates $\grpsu{3}\supset\grpu{1}\times\grpsu{2}$ RCCs by calling the subroutine \texttt{calculate\_coupling\_canonical}.  Internally, this subroutine first calculates the RCCs with the highest or lowest-weight $\epsilon_{3}^{\rm E}\Lambda_{3}^{\rm E}$, depending on whether the desired $\epsilon_{3}$ is closer to the highest or lowest weight, by calling the subroutine \texttt{calculate\_coupling\_canonical\_extremal}. Then the final RCCs are calculated from the extremal RCCs by calling the subroutine \texttt{calculate\_coupling\_canonical\_nonextremal}.

\subsection{Module \texttt{ndsu3lib\_coupling\_su3so3}}
\label{sec:implementation:su3so3}

The user calculates $\grpsu{3}\supset\grpso{3}$ RCCs by calling the subroutine \texttt{calculate\_coupling\_su3so3}. Internally, this subroutine first calls the subroutine \texttt{calculate\_orthonormalization\_matrix} to calculate the orthonormalization matrices $O^{(\lambda,\mu)L}$, and then the subroutine \texttt{calculate\_coupling\_su3so3\_internal} is invoked to calculate the RCCs themselves. Both these subroutines make use of inner products of $\grpsu{3}\supset\grpu{1}\times\grpsu{2}$ and Elliott basis states, given by~(\ref{26}), which are provided by the subroutine \texttt{calculate\_transformation\_coef}.

To avoid loss of precision when evaluating transformation brackets between $\grpsu{3}\supset\grpu{1}\times\grpsu{2}$ and orthonormal $\grpsu{3}\supset\grpso{3}$ bases, the evaluation of Eq.~(\ref{26}) and the orthonormalization~(\ref{ortoc}) can be done with either double or quadruple precision floating-point arithmetic in hardware, or multiprecision floating-point arithmetic in software. The precision is selected internally at run time in a way which was empirically optimized through testing the unitarity of the transformation brackets~(\ref{ortoc}), to avoid usage of unnecessarily high precision, which would increase the computation time.  A heuristic set of criteria are used to select the precision, based on the quantum numbers $\lambda$, $\mu$, and $L$.  While the detailed rules are more complex, and may be found in the code for this module, we note that, for $\lambda+\mu+L\le17$, double precision is always used, while, for $18\le \lambda+\mu+L\le59$, either double or quadruple precision may be used, and, for $\lambda+\mu+L\ge60$, multiprecision precision might also be selected.  For multiprecision arithmetic, \texttt{ndsu3lib} by default requests 37-digit precision from the MPFUN2020 library, providing an incremental but sometimes relevant (Sect.~\ref{sec:precision:physical}) improvement over quadruple precision (approximately 34 digits).  If ever needed for more extreme applications, an increased precision could be selected at compile time by increasing the value of the parameter \texttt{ndig} in the module \texttt{ndsu3lib\_tools} (Sect.~\ref{sec:implementation:tools}) to the desired number of digits.

Since not all compilers or hardware support quadruple precision, and since multiprecision arithmetic requires an external library, the use of quadruple precision or multiprecision arithmetic is optional and must be enabled at compilation.  If both quadruple precision and multiprecision are available, it is typically recommended to enable them, to ensure reliably precise results without unnecessarily increasing the computation time. If quadruple precision is not supported in hardware, the multiprecision library can also be used to emulate quadruple precision, albeit at a cost in performance. The effect of different choices of precision on the calculated results is discussed in Sect.~\ref{sec:precision:physical}.

\subsection{Module \texttt{ndsu3lib\_recoupling}}

The $U$, $Z$, and 9-($\lambda,\mu$) coefficients are calculated by the subroutines \texttt{calculate\_u\_coef}, \texttt{calculate\_z\_coef}, and \texttt{calculate\_9\_lambda\_mu}, respectively. To solve the systems of linear equations~(\ref{22m}) and~(\ref{z}), these subroutines call the subroutine \texttt{dgesv} from the LAPACK library.

 \section{Validation and precision}
\label{sec:precision}

In this sections we describe how we tested the validity of computed RCCs, using the method of infinitesimal generators, and the precision of computed RCCs and recoupling coefficients, using orthonormality relations. The precision of \texttt{ndsu3lib} is compared to the precision of the AD library and \texttt{SU3lib}.

Valid RCCs must satisfy the equations~(\ref{racgen}) and analogous equations for $\grpsu{3}\supset\grpso{3}$ RCCs stemming from the method of infinitesimal generators. These equations provide self-contained tests of validity, which do not require any externally provided benchmark values. We tested the validity of SU(3) RCCs computed by \texttt{ndsu3lib} for a limited set of SU(3) quantum numbers by checking that the RCCs satisfy these equations.

The RCCs and recoupling coefficients must also satisfy orthonormality relations, which provide tests of numerical precision and are less complex than the equations~(\ref{racgen}), allowing tests for a much greater range of quantum numbers. Each orthonormality relation has certain fixed parameters (irrep quantum numbers and branching quantum numbers) and certain summed-over dummy indices (the remaining quantum numbers). We use the orthonormality relations to test \texttt{ndsu3lib} by evaluating the sum and comparing it to 0 or 1. We define the error as the difference between the sum and 0 or 1, whichever is expected.

To see how precision varies as the quantum numbers increase, we develop a systematic series of tests, by taking tests with a certain sum $\Sigma$ of quantum numbers. Due to rapid growth of the time needed to take all the possible tests with increasing $\Sigma$, we do not take all the possible tests for larger values of $\Sigma$ and resort to random sampling as specified later. We then plot the maximal and mean errors as functions of $\Sigma$.

\subsection{$\grpsu{3}\supset\grpu{1}\times\grpsu{2}$ reduced coupling coefficients}
\label{sec:precision:canonical}

Here we check how well the computed $\grpsu{3}\supset\grpu{1}\times\grpsu{2}$ RCCs satisfy the orthonormality relation~(\ref{ortocan}).

\begin{figure}
\includegraphics[width=\linewidth]{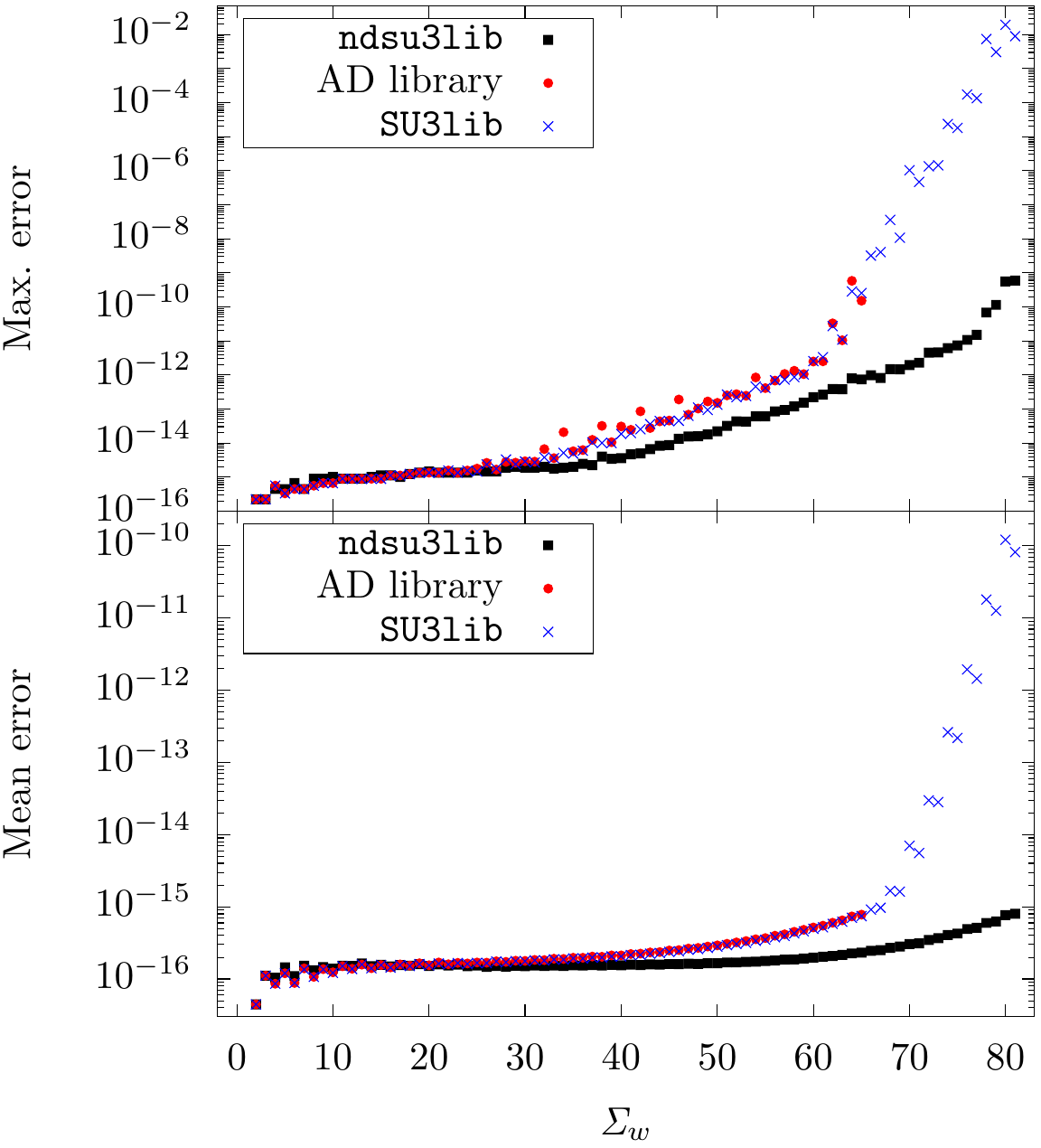}
\caption{The maximal (top) and mean (bottom) errors for $\grpsu{3}\supset\grpu{1}\times\grpsu{2}$ RCCs as functions of $\Sigma_{w}$ for \texttt{ndsu3lib} (squares), the AD library (circles), and \texttt{SU3lib} (crosses) (lower is better).}
\label{wc_error}
\end{figure}

Fig.~\ref{wc_error} shows the maximal and mean errors as functions of $\Sigma_{w}\equiv\lambda_{1}+\mu_{1}+\lambda_{2}+\mu_{2}+\lambda_{3}+\mu_{3}$. The maximal and mean errors tend to increase as the quantum numbers increase, and for \texttt{ndsu3lib} they reach the values of approximately $10^{-9}$ and $10^{-15}$, respectively, for $\Sigma_{w}=81$.

The precisions of the three libraries are comparable for small quantum numbers.  However, with increasing quantum numbers the AD library and \texttt{SU3lib} lose precision more rapidly than \texttt{ndsu3lib}.  Moreover, starting from $\Sigma_{w}=66$, there are examples of RCCs for which the AD library fails outright, by which we mean not simply an incremental loss of precision, but rather an error on the order of unity or greater.  Consequently, no results are shown for the AD library in Fig.~\ref{wc_error} for $\Sigma_{w}\geq66$. In contrast, the orthonormality relation~(\ref{ortocan}) was verified exhaustively for all sets of RCCs with $\Sigma_w\leq81$ for \texttt{ndsu3lib}, with no such failures found, as can be seen from the maximal errors of $\lesssim 10^{-10}$ in Fig.~\ref{wc_error}~(top).

\subsection{$\grpsu{3}\supset\grpso{3}$ reduced coupling coefficients}
\label{sec:precision:physical}

Here we check how well the computed $\grpsu{3}\supset\grpso{3}$ RCCs satisfy the orthonormality relation~(\ref{ortowp}).

\begin{figure}
\includegraphics[width=\linewidth]{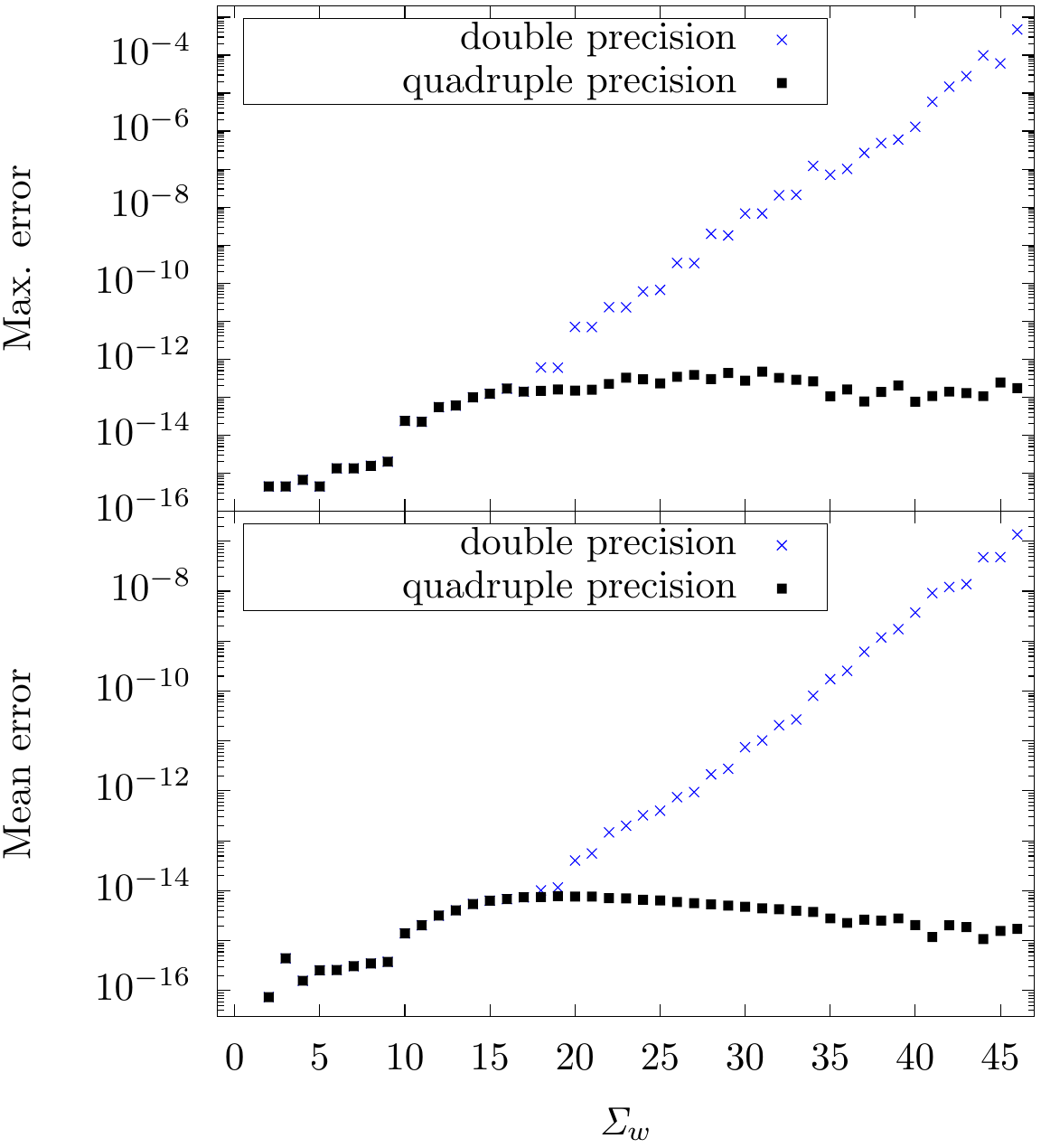}
\caption{The maximal (top) and mean (bottom) errors for $\grpsu{3}\supset\grpso{3}$ RCCs as functions of $\Sigma_{w}$ obtained using only double precision (crosses) and allowing the quadruple precision (squares) (lower is better). }
\label{fig03.pdf}
\end{figure}

Results are affected by the choice of the precision of floating-point calculations. Fig.~\ref{fig03.pdf} shows the maximal and mean errors as functions of $\Sigma_w$, either in the case where arithmetic is restricted to double precision (crosses) or where quadruple precision is also enabled for automatic selection (squares) as described in Sect.~\ref{sec:implementation:su3so3}. The errors of the double-precision computations increase approximately exponetially with increasing quantum numbers, eventually reaching the point where the errors are comparable to the values themselves. In contrast, the errors obtained allowing quadruple precision depend only weakly on the quantum numbers and remain below about $10^{-12}$ over the range explored.

However, for values of the quantum numbers much larger than those explored in Fig.~\ref{fig03.pdf} (and larger than encountered in typical practical applications in nuclear physics), errors obtained using quadruple precision can increase to the point that they might become of concern, and may be improved through the use of multiprecision arithmetic.  For example, for the coupling $(7,39)\times(41,2)\to(3,8)$, the maximal and mean errors obtained using the quadruple precision are approximately $10^{-9}$ and $10^{-10}$, respectively, whereas those obtained allowing multiprecision arithmetic (with 37-digit precision) are approximately $10^{-15}$.  In all subsequent results shown in this work, calculations are carried out with quadruple precision enabled, and the quantum numbers involved are not large enough to trigger multiprecision arithmetic.

Furthermore, WIGXJPF provides more reliable and accurate results than GSL for angular momentum coupling and recoupling coefficients, for large angular momenta~\cite{wigxjpf}. For large quantum numbers, use of WIGXJPF can reduce the errors in the results calculated by \texttt{ndsu3lib} by several orders of magnitude compared to results obtained using GSL. For example, for the coupling $(30,6)\times(5,10)\to(25,3)$, the maximal error obtained using GSL is approximately $10^{-11}$, whereas with WIGXJPF it is approximately $10^{-14}$. Thus, WIGXJPF should be used as the library for angular momentum coupling and recoupling coefficients, if maximal precision is required in calculations involving large values of quantum numbers.  All results shown in the present work are obtained using WIGXJPF.  However, use of GSL does not noticeably degrade the precision for the range of quantum numbers explored in the present study.

\begin{figure}
\includegraphics[width=\linewidth]{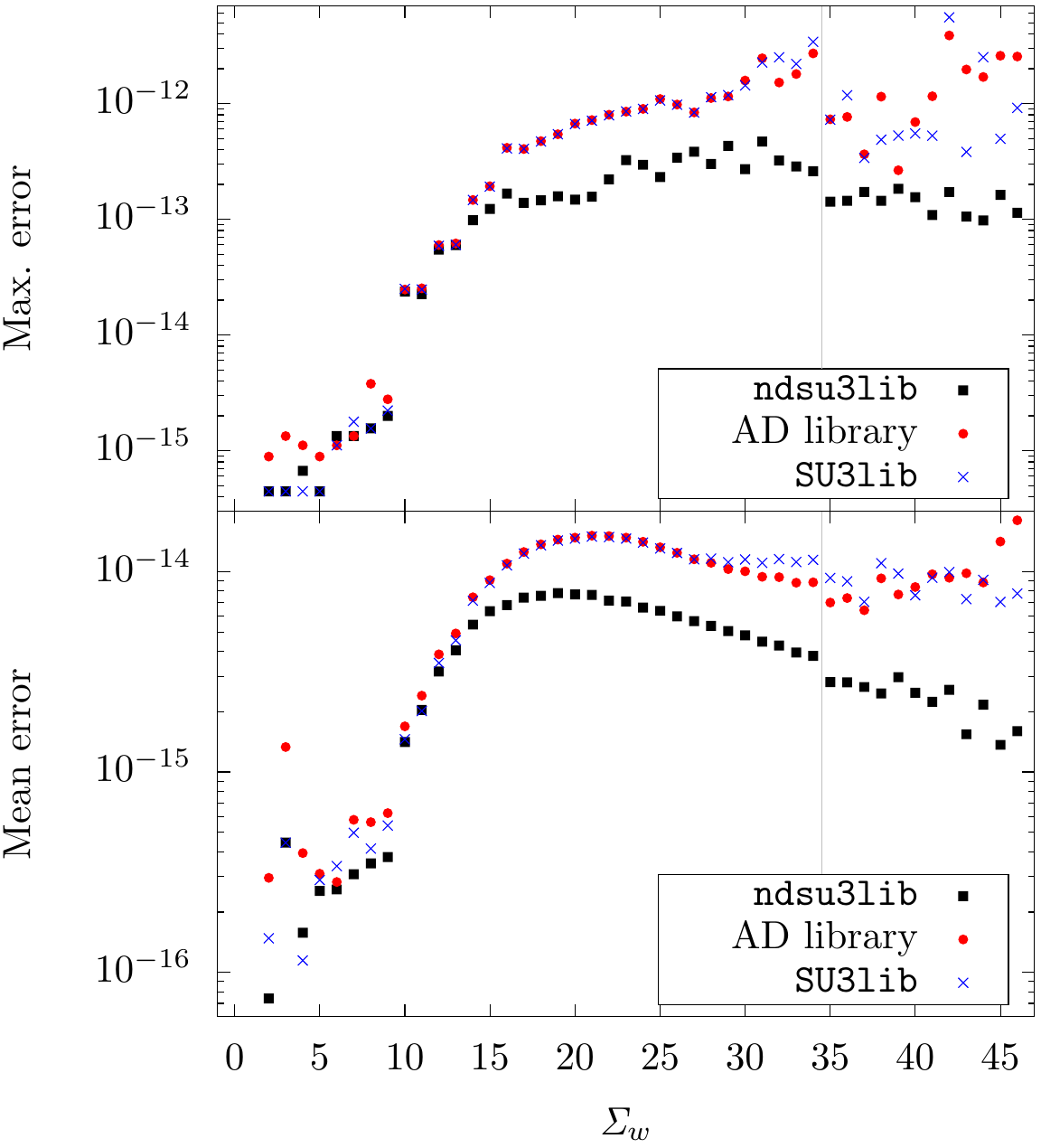}
\caption{The maximal (top) and mean (bottom) errors for $\grpsu{3}\supset\grpso{3}$ RCCs as functions of $\Sigma_{w}$ for \texttt{ndsu3lib} (squares), the AD library (circles), and \texttt{SU3lib} (crosses) (lower is better). For $\Sigma_{w}\ge35$ (to the right of the gray vertical line), computations were made for only 100 randomly selected sets of $\grpsu{3}$ quantum numbers.}
\label{wp_error}
\end{figure}

Fig.~\ref{wp_error} compares the maximal and mean errors of \texttt{ndsu3lib}, the AD library, and \texttt{SU3lib} as functions of $\Sigma_w$ (with quadruple precision arithmetic enabled for these other libraries, as well).  For $\Sigma_{w}\ge35$ (to the right of the gray vertical line), computations were made for only 100 randomly selected sets of $\grpsu{3}$ quantum numbers. With increasing quantum numbers, the errors tend to increase and then saturate (the mean error tends to slightly decrease). When the random sampling starts, the errors decrease little\footnote{The decrease of the maximal error is not surprising, because random sampling is likely to miss extremal cases.} and start exhibiting less systematic behavior. For greater quantum numbers \texttt{ndsu3lib} is slightly more precise than the other two libraries. A systematic comparison for $\Sigma_{w}>46$ was not done due to very long computation time.

\subsection{$U$ recoupling coefficients}
\label{sec:precision:u}

Here we check how well the computed $U$ recoupling coefficients satisfy the orthonormality relation
\begin{multline}\label{ortou}
\sum_{\substack{\lambda_{12}\mu_{12}\\\rho_{12}\rho_{12,3}}}U[(\lambda_{1},\mu_{1})(\lambda_{2},\mu_{2})(\lambda,\mu)(\lambda_{3},\mu_{3});\\
(\lambda_{12},\mu_{12})\rho_{12}\rho_{12,3}(\lambda_{23},\mu_{23})\rho_{23}\rho_{1,23}]\\
\times U[(\lambda_{1},\mu_{1})(\lambda_{2},\mu_{2})(\lambda,\mu)(\lambda_{3},\mu_{3});\\
(\lambda_{12},\mu_{12})\rho_{12}\rho_{12,3}(\lambda_{23}',\mu_{23}')\rho_{23}'\rho_{1,23}']\\
=\delta_{\lambda_{23}\lambda_{23}'}\delta_{\mu_{23}\mu_{23}'}\delta_{\rho_{23}\rho_{23}'}\delta_{\rho_{1,23}\rho_{1,23}'}.
\end{multline}

\begin{figure}
\includegraphics[width=\linewidth]{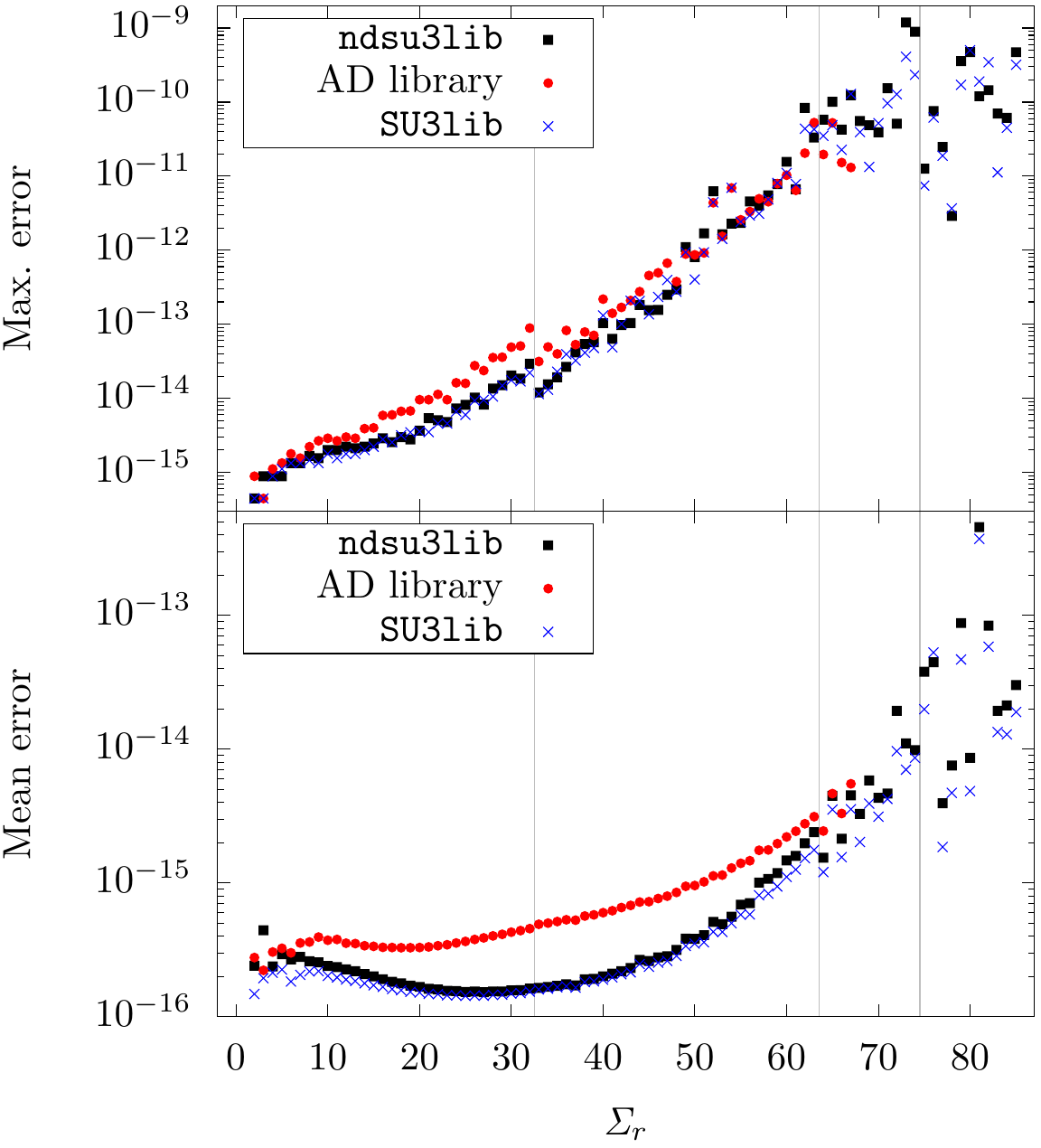}
\caption{The maximal (top) and mean (bottom) errors for $U$ recoupling coefficients as functions of $\Sigma_{r}$ for \texttt{ndsu3lib} (squares), the AD library (circles), and \texttt{SU3lib} (crosses) (lower is better). For $33\le\Sigma_{r}\le63$, computations were made for only 10000 randomly selected sets of the $\grpsu{3}$ quantum numbers in the sum $\Sigma_{r}$; for $64\le\Sigma_{r}\le74$ only 1000 random sets were selected, and for $\Sigma_{r}\ge75$ only 100 random sets were selected. These intervals of $\Sigma_{r}$ are indicated by the gray vertical lines.}
\label{fig05.pdf}
\end{figure}

Fig.~\ref{fig05.pdf} shows the maximal and mean errors as functions of $\Sigma_{r}\equiv\lambda_{1}+\mu_{1}+\lambda_{2}+\mu_{2}+\lambda+\mu+\lambda_{3}+\mu_{3}$. For $33\le\Sigma_{r}\le63$, computations were made for only 10000 randomly selected sets of the $\grpsu{3}$ quantum numbers in the sum $\Sigma_{r}$; for $64\le\Sigma_{r}\le74$ only 1000 random sets were selected, and for $\Sigma_{r}\ge75$ only 100 random sets were selected. (These intervals of $\Sigma_{r}$ are indicated by gray vertical lines.) The maximal error tends to increase as the quantum numbers increase and reaches the value of approximately $10^{-9}$ for $\Sigma_{r}\approx80$. As the quantum numbers increase, the mean error tends to increase in the region of small quantum numbers, then decrease little (like for the $\grpsu{3}\supset\grpso{3}$ RCCs in Fig.~\ref{wp_error}), and then increase. It reaches the value of approximately $10^{-13}$ for $\Sigma_{r}\approx80$. We can see a small decrease of the maximal error when the random sampling starts. For 1000 or less random samples the errors exhibit less systematic behavior.

The precisions of the three libraries are comparable. However, starting from $\Sigma_{r}=68$, the AD library produces incorrect results, which is indicated by missing data in Fig.~\ref{fig05.pdf}.

\subsection{$Z$ recoupling coefficients}
\label{sec:precision:z}

Here we check how well the computed $Z$ recoupling coefficients satisfy the orthonormality relation
\begin{multline}\label{ortoz}
\sum_{\substack{\lambda_{12}\mu_{12}\\\rho_{12}\rho_{12,3}}}Z[(\lambda_{2},\mu_{2})(\lambda_{1},\mu_{1})(\lambda,\mu)(\lambda_{3},\mu_{3});\\
(\lambda_{12},\mu_{12})\rho_{12}\rho_{12,3}(\lambda_{13},\mu_{13})\rho_{13}\rho_{13,2}]\\
\times Z[(\lambda_{2},\mu_{2})(\lambda_{1},\mu_{1})(\lambda,\mu)(\lambda_{3},\mu_{3});\\
(\lambda_{12},\mu_{12})\rho_{12}\rho_{12,3}(\lambda_{13}',\mu_{13}')\rho_{13}'\rho_{13,2}']\\
=\delta_{\lambda_{13}\lambda_{13}'}\delta_{\mu_{13}\mu_{13}'}\delta_{\rho_{13}\rho_{13}'}\delta_{\rho_{13,2}\rho_{13,2}'}.
\end{multline}

\begin{figure}
\includegraphics[width=\linewidth]{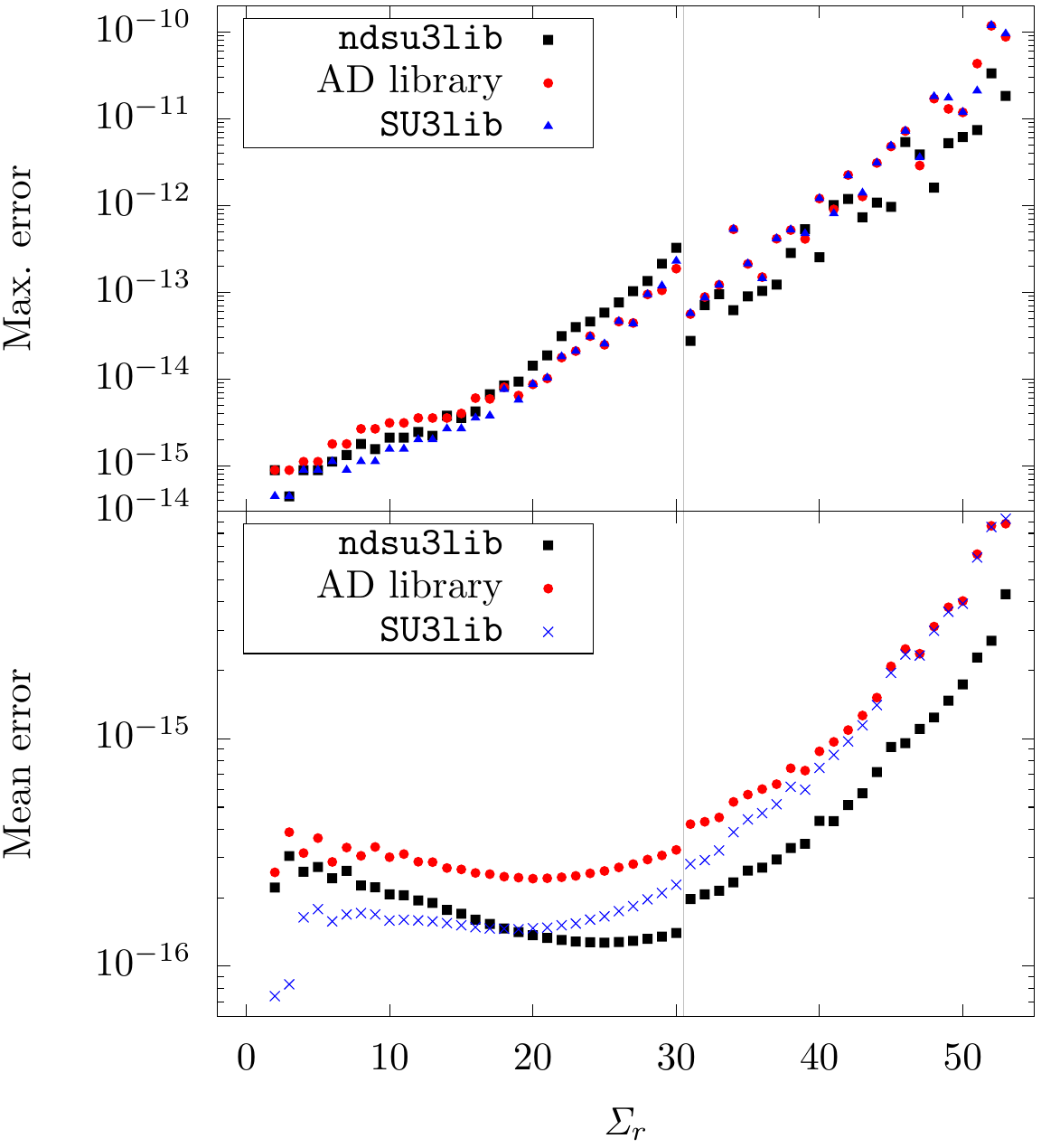}
\caption{The maximal (top) and mean (bottom) errors for $Z$ recoupling coefficients as functions of $\Sigma_{r}$ for \texttt{ndsu3lib} (squares), the AD library (circles), and \texttt{SU3lib} (crosses) (lower is better). For $\Sigma_{r}\ge31$ (to the right of the gray vertical line), computations were made for only 10000 randomly selected sets of the $\grpsu{3}$ quantum numbers in the sum $\Sigma_{r}$.}
\label{fig06.pdf}
\end{figure}

Fig.~\ref{fig06.pdf} shows the maximal and mean errors as functions of $\Sigma_{r}$. For $\Sigma_{r}\ge31$ (to the right of the gray vertical line), computations were made for only 10000 randomly selected sets of the $\grpsu{3}$ quantum numbers in the sum $\Sigma_{r}$. The maximal error tends to increase as the quantum numbers increase and reaches the value of approximately $10^{-10}$ for $\Sigma_{r}=53$. As the quantum numbers increase, the mean error tends to increase in the region of small quantum numbers, then decrease little, and then increase (like for the $U$ recoupling coefficients in Fig.~\ref{fig05.pdf}). It reaches the value of approximately $10^{-14}$ for $\Sigma_{r}=53$. We can see a little decrease (increase) of the maximal (mean) error when the random sampling starts. The precisions of the three libraries are comparable.

\subsection{9-($\lambda,\mu$) coefficients}
\label{sec:precision:9lm}

Here we check how well the computed 9-($\lambda,\mu$) coefficients satisfy the orthonormality relation
\begin{multline}\label{orto9}
\sum_{\substack{\lambda_{13}\mu_{13}\lambda_{24}\mu_{24}\\\rho_{13}\rho_{24}\rho_{13,24}}}\left[\begin{array}{cccc}
(\lambda_{1},\mu_{1}) & (\lambda_{2},\mu_{2}) & (\lambda_{12},\mu_{12}) & \rho_{12} \\
(\lambda_{3},\mu_{3}) & (\lambda_{4},\mu_{4}) & (\lambda_{34},\mu_{34}) & \rho_{34} \\
(\lambda_{13},\mu_{13}) & (\lambda_{24},\mu_{24}) & (\lambda,\mu) & \rho_{13,24} \\
\rho_{13} & \rho_{24} & \rho_{12,34} &  \\
\end{array}\right]\\
\times\left[\begin{array}{cccc}
(\lambda_{1},\mu_{1}) & (\lambda_{2},\mu_{2}) & (\lambda_{12}',\mu_{12}') & \rho_{12}' \\
(\lambda_{3},\mu_{3}) & (\lambda_{4},\mu_{4}) & (\lambda_{34}',\mu_{34}') & \rho_{34}' \\
(\lambda_{13},\mu_{13}) & (\lambda_{24},\mu_{24}) & (\lambda,\mu) & \rho_{13,24} \\
\rho_{13} & \rho_{24} & \rho_{12,34}' &  \\
\end{array}\right]\\
=\delta_{\rho_{12}\rho_{12}'}\delta_{\lambda_{12}\lambda_{12}'}\delta_{\mu_{12}\mu_{12}'}\delta_{\rho_{34}\rho_{34}'}\delta_{\lambda_{34}\lambda_{34}'}\delta_{\mu_{34}\mu_{34}'}\delta_{\rho_{12,34}\rho_{12,34}'}.
\end{multline}
Only results obtained with \texttt{SU3lib} are shown for comparison.

\begin{figure}
\includegraphics[width=\linewidth]{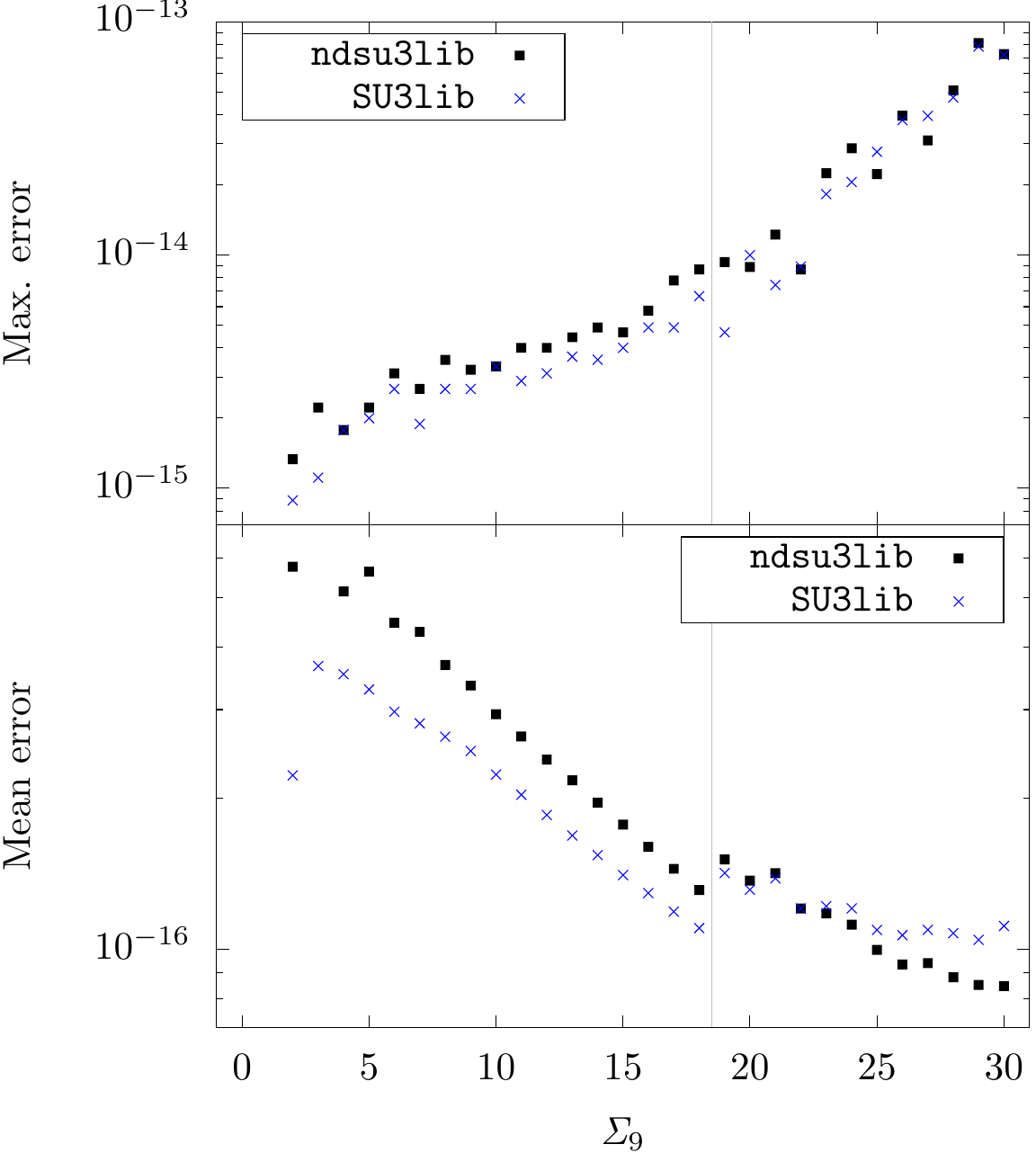}
\caption{The maximal (top) and mean (bottom) errors for 9-($\lambda,\mu$) coefficients as functions of $\Sigma_{9}$ for \texttt{ndsu3lib} (squares) and \texttt{SU3lib} (crosses) (lower is better). For $\Sigma_{9}\ge19$ (to the right of the gray vertical line), computations were made for only 10000 randomly selected sets of the $\grpsu{3}$ quantum numbers in the sum $\Sigma_{9}$.}
\label{fig07.pdf}
\end{figure}

Fig.~\ref{fig07.pdf} shows the maximal and mean errors as functions of $\Sigma_{9}\equiv\lambda_{1}+\mu_{1}+\lambda_{2}+\mu_{2}+\lambda_{3}+\mu_{3}+\lambda_{4}+\mu_{4}+\lambda+\mu$. For $\Sigma_{9}\ge19$ (to the right of the gray vertical line), computations were made for only 10000 randomly selected sets of the $\grpsu{3}$ quantum numbers in the sum $\Sigma_{9}$. The maximal error tends to increase as the quantum numbers increase and reaches the value of approximately $10^{-13}$ for $\Sigma_{r}=30$. The mean error tends to decrease in this limited range of quantum numbers with values around $10^{-16}$, exhibiting little jump when the random sampling starts. However, an indication of a stop of the decrese for the largest quantum numbers can be observed (a similar behavior was observed for the $U$ and $Z$ recoupling coefficients in Figs.~\ref{fig05.pdf} and~\ref{fig06.pdf}). The precisions of the two libraries are comparable.

 \section{Speed}
\label{sec:speed}

To investigate the performance of \texttt{ndsu3lib} we measured how much time the computation of RCCs and recoupling coefficients takes. Results obtained using the AD library and \texttt{SU3lib} are included as well for comparison. The results in this section were obtained by serial computation using the Intel\textsuperscript{\textregistered} Xeon\textsuperscript{\textregistered} CPU E5-2680 v3 with clock speed of 2.50 GHz, the GNU Compiler Collection with the \texttt{O3} optimization level, the Intel\textsuperscript{\textregistered} Math Kernel Library for the LAPACK subroutine solving systems of linear equations, and the WIGXJPF library for angular momentum coupling and recoupling coefficients.

\subsection{$\grpsu{3}\supset\grpu{1}\times\grpsu{2}$ reduced coupling coefficients}
\label{sec:speed:canonical}

\begin{figure}
\includegraphics[width=\linewidth]{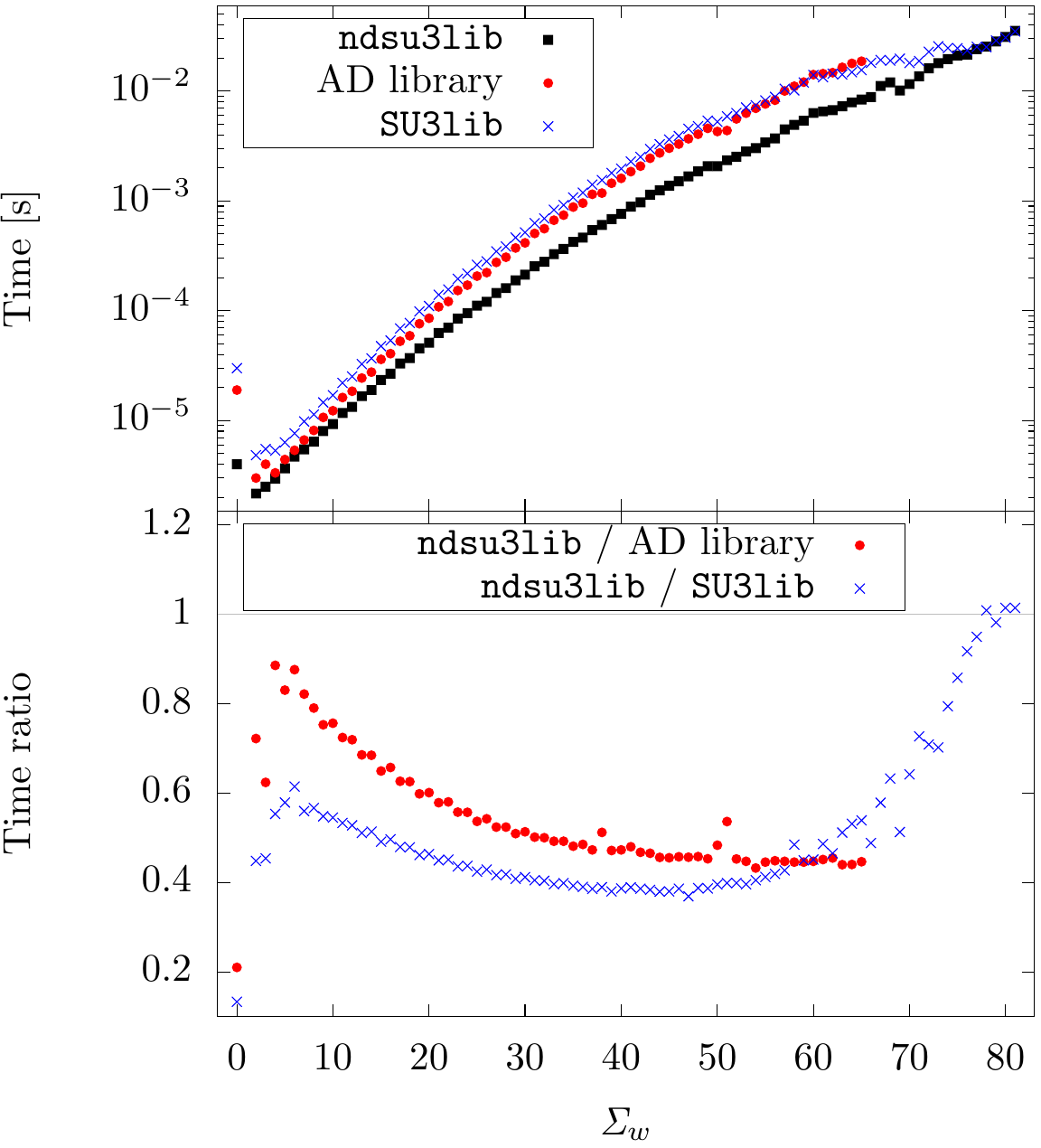}
\caption{Time spent calculating $\grpsu{3}\supset\grpu{1}\times\grpsu{2}$ RCCs divided by the number of $\grpsu{3}$ couplings, as a function of $\Sigma_{w}$, for \texttt{ndsu3lib} (squares), the AD library (circles), and \texttt{SU3lib} (crosses) (lower is better). Ratios of the time spent by \texttt{ndsu3lib} over the times spent by the AD library (circles) and \texttt{SU3lib} (crosses) are shown as well (lower is better for \texttt{ndsu3lib}).}
\label{wc_time}
\end{figure}

Fig.~\ref{wc_time} shows the time spent computing all the $\grpsu{3}\supset\grpu{1}\times\grpsu{2}$ RCCs divided by the number of possible $\grpsu{3}$ couplings as a function of $\Sigma_{w}$. The figure also shows the ratios of the time spent by \texttt{ndsu3lib} over the times spent by the AD library and \texttt{SU3lib}. The data obtained with the AD library for $\Sigma_{w}>65$ are missing, because the library produces incorrect results for such $\Sigma_{w}$. Our library is faster than the other 2 libraries by a factor which slowly increses with increasing $\Sigma_{w}$ and reaches the value of approximately two for $\Sigma_{w}=65$. As $\Sigma_{w}$ increases beyond the value of 65, the ratio of the time spent by \texttt{ndsu3lib} over the time spent by \texttt{SU3lib} increases. However, for such $\Sigma_{w}$ the error of \texttt{SU3lib} increases with increasing $\Sigma_{w}$ faster than the error of \texttt{ndsu3lib} as shown in Fig.~\ref{wc_error}.

\subsection{$\grpsu{3}\supset\grpso{3}$ coupling coefficients}
\label{sec:speed:physical}

\begin{figure}
\includegraphics[width=\linewidth]{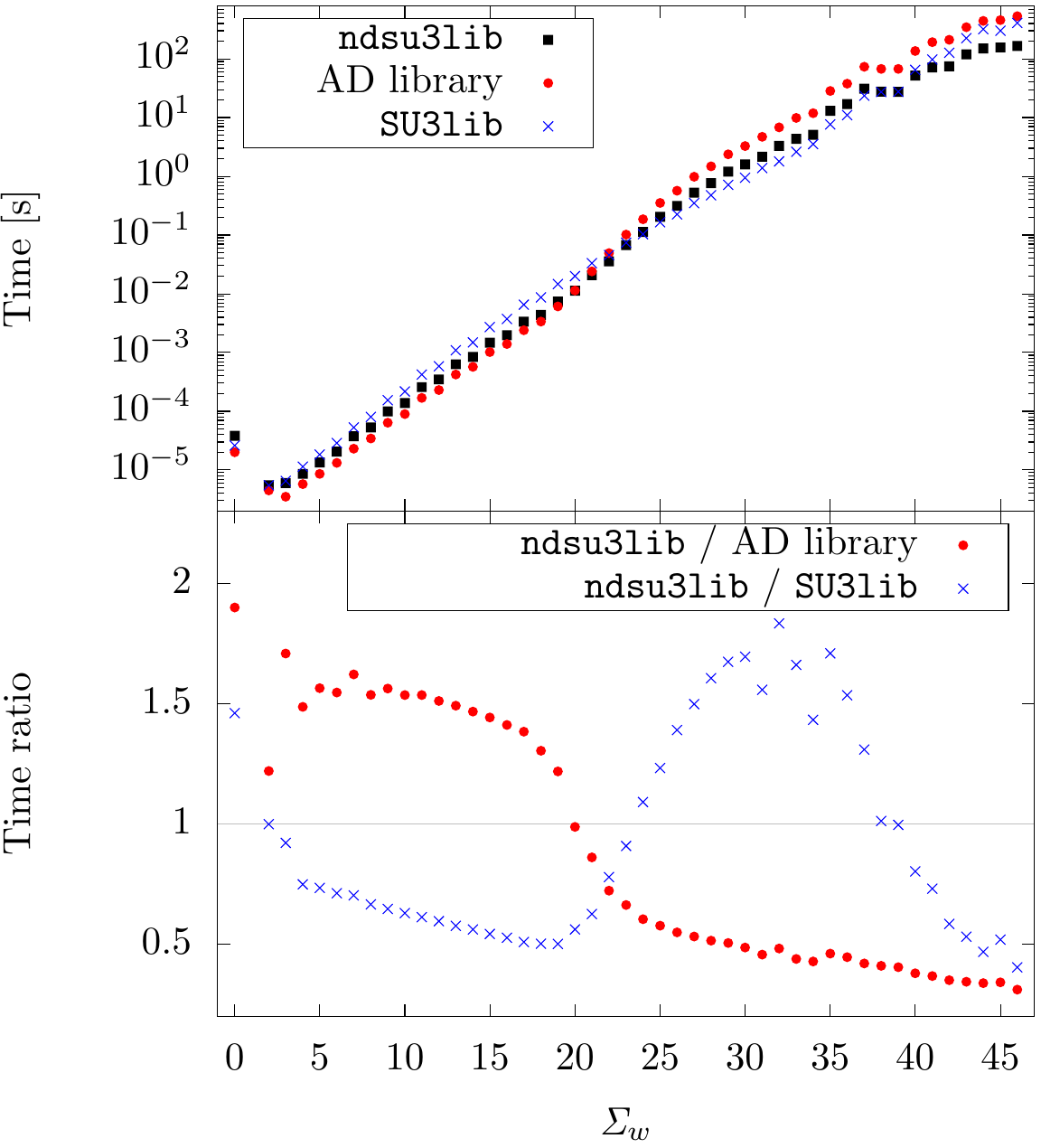}
\caption{Time spent calculating $\grpsu{3}\supset\grpso{3}$ RCCs, presented as described in the caption to Fig.~\ref{wc_time}.}
\label{wp_time}
\end{figure}

Fig.~\ref{wp_time} shows the time spent computing the $\grpsu{3}\supset\grpso{3}$ RCCs divided by the number of $\grpsu{3}$ couplings as a function of $\Sigma_{w}$. The figure also shows the ratios of the time spent by \texttt{ndsu3lib} over the times spent by the AD library and \texttt{SU3lib}. The results were obtained allowing quadruple precision for floating-point calculations (the quantum numbers involved are not large enough to trigger multiprecision arithmetic) and disabling caching of inner products of $\grpsu{3}\supset\grpu{1}\times\grpsu{2}$ and Elliott basis states in \texttt{SU3lib}. Starting from $\Sigma_{w}=35$, computations were made for only 100 randomly selected sets of $\grpsu{3}$ quantum numbers.

For $\Sigma_{w}<20$ \texttt{ndsu3lib} is slower than the AD library by a factor which tends to decrease with increasing $\Sigma_{w}$ in the range between 1.2 and 1.9. For $\Sigma_{w}>20$ \texttt{ndsu3lib} is faster than the AD library by a factor which tends to increase with increasing $\Sigma_{w}$ and reaches the value of approximately 3 for $\Sigma_{w}=46$. The comparison between \texttt{ndsu3lib} and \texttt{SU3lib} is different. For $\Sigma_{w}<24$ \texttt{ndsu3lib} is faster than \texttt{SU3lib} by a factor varying in the range between 1 and 2. For $24\le\Sigma_{w}\le37$ \texttt{SU3lib} is faster by a factor varying between 1.1 and 1.8. For $\Sigma_{w}>39$ \texttt{ndsu3lib} is faster by a factor which tends to increase with increasing $\Sigma_{w}$ and reaches the value of approximately 2.5 for $\Sigma_{w}=46$.

\subsection{$U$ recoupling coefficients}
\label{sec:speed:u}

\begin{figure}
\includegraphics[width=\linewidth]{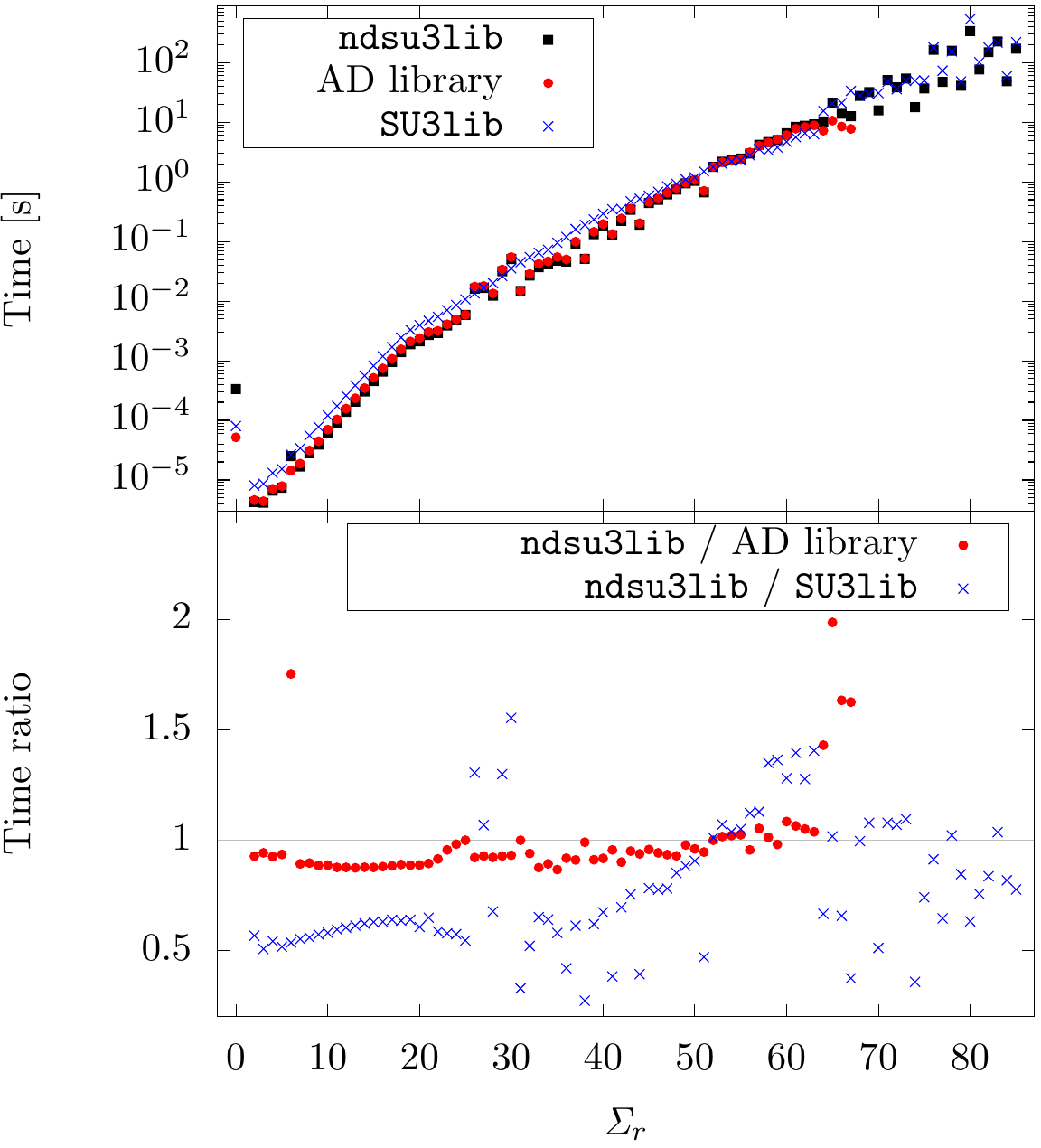}
\caption{Time spent calculating $U$ recoupling coefficients divided by the number of sets of the $\grpsu{3}$ quantum numbers for the given value of $\Sigma_r$, as a function of $\Sigma_{r}$, for \texttt{ndsu3lib} (squares), the AD library (circles), and \texttt{SU3lib} (crosses) (lower is better). Ratios of the time spent by \texttt{ndsu3lib} over the times spent by the AD library (circles) and \texttt{SU3lib} (crosses) are shown as well (lower is better for \texttt{ndsu3lib}).}
\label{fig10.pdf}
\end{figure}

Fig.~\ref{fig10.pdf} shows the time spent computing the $U$ recoupling coefficients divided by the number of sets of the $\grpsu{3}$ quantum numbers in the sum $\Sigma_{r}$ as a function of $\Sigma_{r}$. The figure also shows the ratios of the time spent by \texttt{ndsu3lib} over the times spent by the AD library and \texttt{SU3lib}. Starting from $\Sigma_{r}=33$, computations were made for only a limited number of randomly selected sets of the $\grpsu{3}$ quantum numbers in the sum $\Sigma_{r}$ as described in Sect.~\ref{sec:precision:u}. For $\Sigma_{r}=6$, \texttt{ndsu3lib} is slower than the AD library by a factor of 1.75. Apart from this case, the average speed of both libraries is comparable up to $\Sigma_r=63$. For $64\le\Sigma_r\le67$ the AD library is faster by a factor ranging between 1.4 and 2, and for $\Sigma_r\ge68$ the AD library starts producing incorrect results, which is indicated by missing data in Fig.~\ref{fig10.pdf}. The ratio of the time spent by \texttt{ndsu3lib} over the time spent by \texttt{SU3lib} does not exhibit a specific pattern and is scattered in the range between 0.25 and 1.6.

\subsection{$Z$ recoupling coefficients}
\label{sec:speed:z}

\begin{figure}
\includegraphics[width=\linewidth]{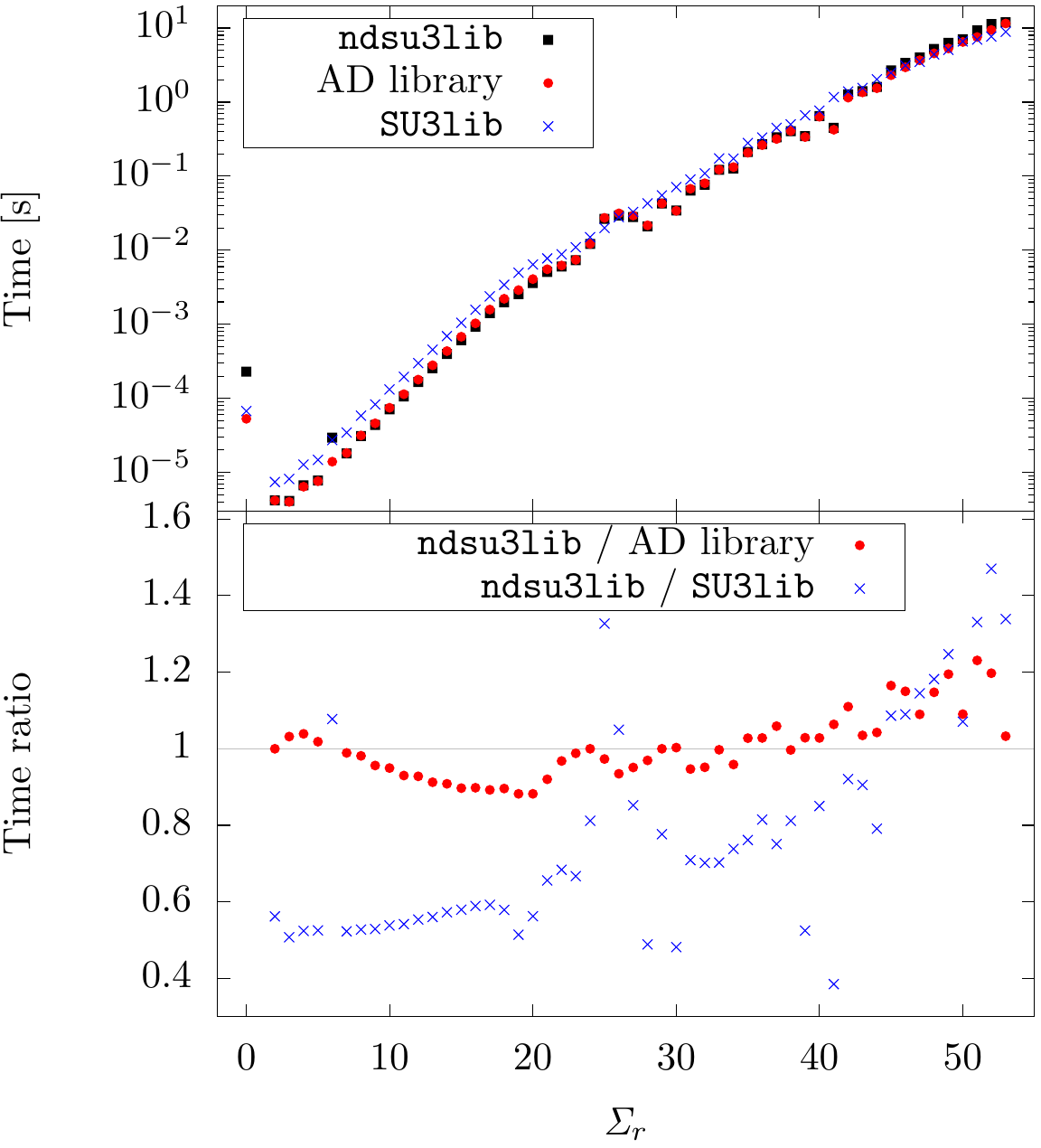}
\caption{Time spent calculating $Z$ recoupling coefficients, presented as described in the caption to Fig.~\ref{fig10.pdf}.}
\label{fig11.pdf}
\end{figure}

Fig.~\ref{fig11.pdf} shows the time spent computing the $Z$ recoupling coefficients divided by the number of sets of the $\grpsu{3}$ quantum numbers in the sum $\Sigma_{r}$ as a function of $\Sigma_{r}$. The figure also shows the ratios of the time spent by \texttt{ndsu3lib} over the times spent by the AD library and \texttt{SU3lib}. Starting from $\Sigma_{r}=31$, computations were made for only 10000 randomly selected sets of the $\grpsu{3}$ quantum numbers in the sum $\Sigma_{r}$. The average speeds of \texttt{ndsu3lib} and the AD library are comparable. The ratio of the time spent by \texttt{ndsu3lib} over the time spent by \texttt{SU3lib} does not exhibit a specific pattern and is scattered in the range between 0.4 and 1.5.

\subsection{9-($\lambda,\mu$) coefficients}
\label{sec:speed:9lm}

In this section only results obtained with \texttt{SU3lib} are shown for comparison.

\begin{figure}
\includegraphics[width=\linewidth]{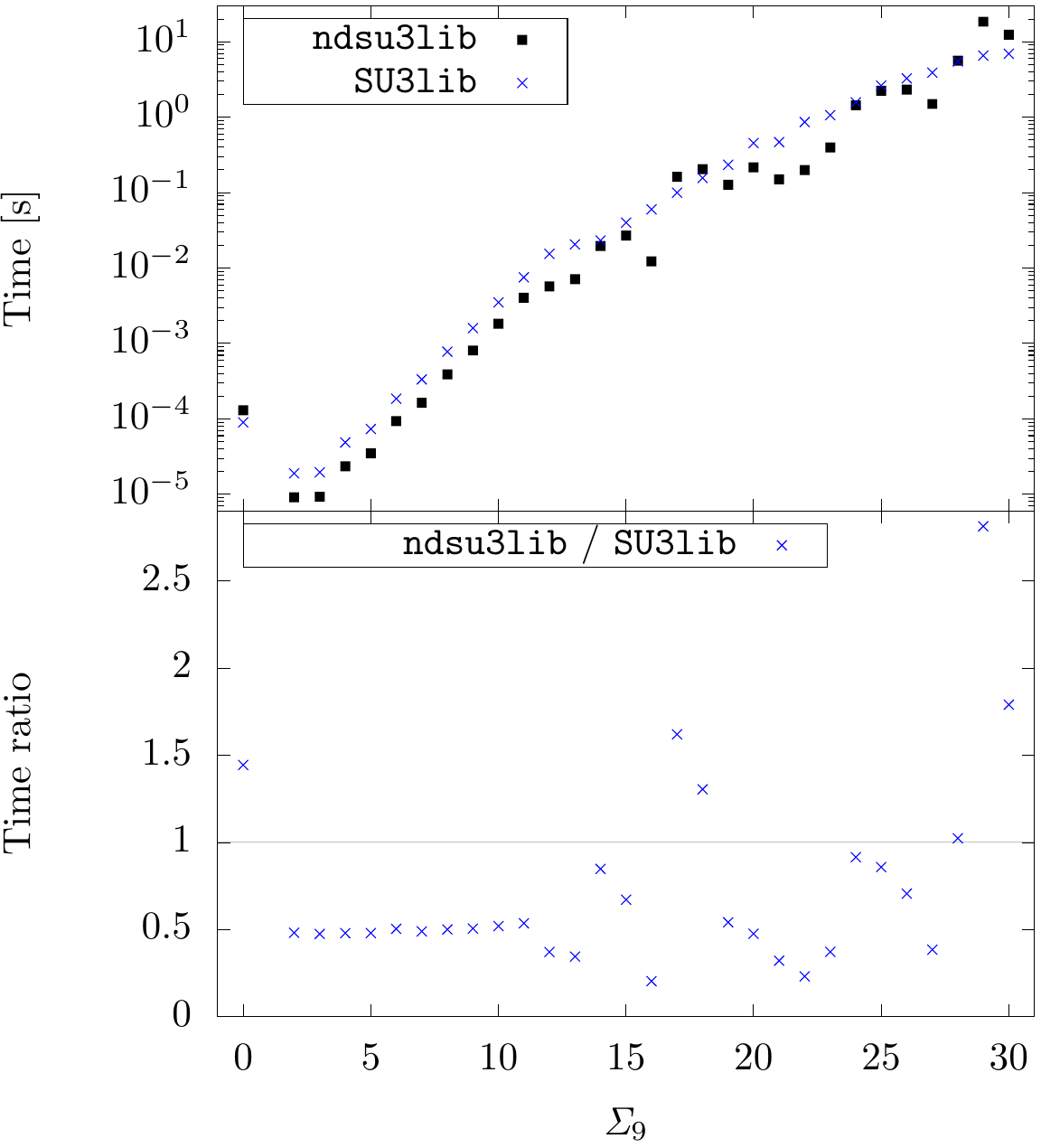}
\caption{Time spent calculating 9-($\lambda,\mu$) coefficients divided by the number of sets of the $\grpsu{3}$ quantum numbers for the given value of $\Sigma_{9}$, as a function of $\Sigma_{9}$, for \texttt{ndsu3lib} (squares) and \texttt{SU3lib} (crosses) (lower is better). Ratio of the time spent by \texttt{ndsu3lib} and \texttt{SU3lib} is shown as well (lower is better for \texttt{ndsu3lib}).}
\label{fig12.pdf}
\end{figure}

Fig.~\ref{fig12.pdf} shows the time spent computing the 9-($\lambda,\mu$) coefficients divided by the number of sets of the $\grpsu{3}$ quantum numbers in the sum $\Sigma_{9}$ as a function of $\Sigma_{9}$. The figure also shows the ratio of the times spent by \texttt{ndsu3lib} and \texttt{SU3lib}. Starting from $\Sigma_{r}=19$, computations were made for only 10000 randomly selected sets of the $\grpsu{3}$ quantum numbers in the sum $\Sigma_{9}$. The ratio of the times spent by \texttt{ndsu3lib} and \texttt{SU3lib} does not exhibit a specific pattern and is scattered in the range between 0.2 and 2.8.

 \section{Conclusion}
\label{sec:conclusion}

A library \texttt{ndsu3lib} for computation of $\grpsu{3}$ reduced coupling coefficients (RCCs) and recoupling coefficients to be used in, \textit{e.g.}, modern \textit{ab initio} nuclear structure calculations in symmetry-guided frameworks, such as the symplectic no-core configuration interaction framework, has been developed.

The library implements the Draayer-Akiyama (DA) algorithms and Millener's algorithms. We provide a self-contained derivation of the DA building-up process for canonical RCCs from a few basic identities for $\grpsu{3}$-reduced matrix elementss, RCCs, and recoupling coefficients, together with the constraints (vanishing conditions) imposed by the Biedenharn-Louck-Hecht resolution of the outer multiplicity problem. We also document the implemented formulae, with minor corrections to expressions in the literature (see~\ref{appendix}).

The DA algorithm is implemented with one improvement: the $\grpsu{3}\supset\grpu{1}\times\grpsu{2}$ RCCs with non-extremal $\epsilon_3\Lambda_3$ are calculated iteratively from those with the lowest-weight $\epsilon_3\Lambda_3$, if the desired $\epsilon_3$ is closer to the lowest weight (see Sect.~\ref{sec:algorithm:canonical} for details). In this way, the number of iterations is reduced, reducing loss of precision (see Fig.~\ref{wc_error}) and computation time (see Fig.~\ref{wc_time}).

To increase the range of quantum numbers for which valid and precise $\grpsu{3}\supset\grpso{3}$ RCCs can be obtained, the calculation of the trasformation brackets between the $\grpsu{3}\supset\grpu{1}\times\grpsu{2}$ and $\grpsu{3}\supset\grpso{3}$ bases can be done with double or quadruple precision or multiprecision floating-point arithmetic. The precision is selected internally at run time in a way which was empirically optimized through testing to avoid usage of unnecessarily high precision, which would increase the computation time.

The algorithms were implemented in an older Akiyama-Draayer (AD) library written in Fortran as well as in a recent C++ library \texttt{SU3lib}, which also provides for OpenMP multithreaded operation and supports the use of multiprecision arithmetic. We compare the performances of these libraries and \texttt{ndsu3lib}.

Some limitations of the AD library have been overcome. In particular, \texttt{ndsu3lib} provides valid results for a larger range of $\grpsu{3}$ quantum numbers.  Furthermore, \texttt{ndsu3lib} makes use of allocatable arrays, so that hard-coded limits are not placed on the set of coefficients which can be evaluated, and it is written in a modern programming language allowing for optimization for modern computer architectures.

Our library provides more accurate $\grpsu{3}\supset\grpu{1}\times\grpsu{2}$ RCCs with large quantum numbers than the AD library and \texttt{SU3lib}.  Moreover, when used in conjunction with multiprecision arithmetic and with the WIGXJPF library for angular momentum coupling coefficient, it provides more accurate $\grpsu{3}\supset\grpso{3}$ RCCs, a case of particular interest in nuclear physics, at larger values for the quantum numbers than the AD library.  For the recoupling coefficients, the precisions of the three libraries are similar. The speeds of the three libraries are comparable.

\begin{acknowledgements}
We thank Tom\'{a}\v{s} Dytrych for valuable discussions.  This material is based
upon work supported by the U.S.~Department of Energy, Office of Science, under
Awards No.~DE-FG02-95ER40934, DE-FG02-00ER41132, and DE-AC02-06CH11357, and by
the 2020 DOE Early Career Award Program.  TRIUMF receives federal funding via a
contribution agreement with the National Research Council of Canada.  This
research used computational resources of the University of Notre Dame Center for
Research Computing, the Digital Research Alliance of Canada, and of the National Energy
Research Scientific Computing Center (NERSC), a DOE Office of Science User
Facility supported by the Office of Science of the U.S.~Department of Energy
under Contract No.~DE-AC02-05CH11231, using NERSC award NP-ERCAP0031993.
\end{acknowledgements}

\appendix

\section{Auxiliary formulae}\label{appendix}

\subsection{$\grpsu{3}\supset\grpso{3}$ basis states}\label{appendix:basis}

The inner multiplicity of a given $L$ within a given $\grpsu{3}$ irrep ($\lambda,\mu$) is given by~\cite{racah-kappamax,jmp-6-1965-142-Kaufman}
\begin{multline}\label{kappamax}
\kappa_{\rm max}=\textrm{max}\left(0,\left\lfloor\frac{\lambda+\mu+2-L}{2}\right\rfloor\right)\\
-\textrm{max}\left(0,\left\lfloor\frac{\lambda\!+\!1\!-\!L}{2}\right\rfloor\right)\!-\!\textrm{max}\left(0,\left\lfloor\frac{\mu\!+\!1\!-\!L}{2}\right\rfloor\right),
\end{multline}
where $\lfloor x\rfloor$ denotes the integer part of $x$.

From~(\ref{so3branch}) it follows that within a given $\grpsu{3}$ irrep ($\lambda,\mu$) the possible values of $K$ for a given $L$ are
\begin{align}
\nonumber K&=K_{\rm min},K_{\rm min}+2,\ldots,K_{\rm min}+2(\kappa_{\rm max}-1)\\
&=K_{1},K_{2},\ldots,K_{\kappa_{\rm max}},
\end{align}
where the minimal value of $K$ can be determined as
\begin{equation}\label{kmin}
K_{\rm min}=\left\{\begin{array}{ll}
f(\lambda,\mu,L), & \quad\lambda<\mu, \\
f(\mu,\lambda,L), & \quad\lambda\ge\mu, \\
\end{array}\right.
\end{equation}
where, in turn,
\begin{equation}
f(\lambda,\mu,L)=\left\{\begin{array}{ll}
g(\lambda,\mu,L), & \quad g(\lambda,\mu,L)\ne0, \\
2\textrm{mod}_{2}(L+\mu), & \quad g(\lambda,\mu,L)=0, \\
\end{array}\right.
\end{equation}
and
\begin{equation}
g(\lambda,\mu,L)\!\!=\!\textrm{max}(0,\!L-\mu)+\textrm{mod}_{2}[\textrm{max}(0,\!L-\mu)+\lambda],
\end{equation}
and $\textrm{mod}_{2}(x)$ is the remainder after division of $x$ by 2.

The orthonormalization matrix $O^{(\lambda,\mu)L}$ appearing in~(\ref{ortos}) is defined recursively by\footnote{The expressions~(\ref{6a})--(\ref{6c}) correspond to~(6a)--(6c) of Ref~\cite{draayer}, where~(6b) of~Ref.~\cite{draayer} contains an exponent of $1/2$ which should not be there.}
\begin{align}
\nonumber O^{(\lambda,\mu)L}_{ii}=&\bigg(\overlapjh{(\lambda,\mu)}{K_{i}LM}{(\lambda,\mu)}{K_{i}LM}\\
&-\sum_{j<i}O^{(\lambda,\mu)L}_{ji}O^{(\lambda,\mu)L}_{ji}\bigg)^{-\frac{1}{2}},\label{6a}\\
\nonumber O^{(\lambda,\mu)L}_{ji}=&O^{(\lambda,\mu)L}_{jj}\bigg(\overlapjh{(\lambda,\mu)}{K_{j}LM}{(\lambda,\mu)}{K_{i}LM}\\
&-\sum_{k<j}O^{(\lambda,\mu)L}_{kj}O^{(\lambda,\mu)L}_{ki}\bigg),\label{6b}\\
O^{(\lambda,\mu)L}_{ij}=&-O^{(\lambda,\mu)L}_{ii}\sum_{j\le k<i}O^{(\lambda,\mu)L}_{kj}O^{(\lambda,\mu)L}_{ki},\label{6c}
\end{align}
where $j<i$, and~\cite{draayer}
\begin{equation}\label{6d}
\overlapjh{(\lambda,\mu)}{K_{j}LM}{(\lambda,\mu)}{K_{i}LM}=\overlapjh{(\lambda,\mu)}{\epsilon^{\rm E}\Lambda^{\rm E}M_{\Lambda}^{\rm E}}{(\lambda,\mu)}{K_{i}LK_{j}},
\end{equation}
where the extremal state is given by~(\ref{e}).

\subsection{Inner products of $\grpsu{3}\supset\grpu{1}\times\grpsu{2}$ and $\grpsu{3}\supset\grpso{3}$ basis states}

If the Elliott basis state is projected from the highest-weight $\grpsu{3}\supset\grpu{1}\times\grpsu{2}$ basis state, the inner product of $\grpsu{3}\supset\grpu{1}\times\grpsu{2}$ and Elliott basis states is given by\footnote{The relation~(\ref{26}) corresponds to~(26) of Ref.~\cite{draayer}, where, in the factor $S_{1}(N_{\Lambda}\Lambda M_{\Lambda}=\Lambda M)$ appearing in the initial equation for the overlap, the arguments $N_{\Lambda}$ and $M_{\Lambda}=\Lambda$ should be interchanged, and, in the expression for $C$, the factor $2L+1$ should not be squared.}
\begin{multline}\label{26}
\overlapjh{(\lambda\mu)}{\epsilon\Lambda M_{\Lambda}}{(\lambda\mu)}{KLM}=(-1)^{\frac{\lambda+K}{2}+L-p}\frac{2L+1}{4^{p}}\\
\times\sqrt{\frac{\left(\!\!\begin{array}{c}
\lambda \\
p \\
\end{array}\!\!\right)\left(\!\!\begin{array}{c}
\mu \\
q \\
\end{array}\!\!\right)\left(\!\!\begin{array}{c}
\lambda+\mu+1 \\
q \\
\end{array}\!\!\right)\left(\!\!\begin{array}{c}
2L \\
L-K \\
\end{array}\!\!\right)}{\left(\!\!\begin{array}{c}
2L \\
L-M \\
\end{array}\!\!\right)\left(\!\!\begin{array}{c}
2\Lambda \\
\Lambda+M_{\Lambda} \\
\end{array}\!\!\right)\left(\!\!\begin{array}{c}
p+\mu+1 \\
q \\
\end{array}\!\!\right)}}\\
\times\sum_{\gamma=0}^{p}\left(\!\!\begin{array}{c}
p \\
\gamma \\
\end{array}\!\!\right)\sum_{\alpha}\left(\!\!\begin{array}{c}
2\Lambda-p+\gamma \\
\alpha \\
\end{array}\!\!\right)\left(\!\!\begin{array}{c}
p-\gamma \\
\Lambda-M_{\Lambda}-\alpha \\
\end{array}\!\!\right)\\
\times I\bigg(\!2\alpha\!+\!M_{\Lambda}\!+\!p\!-\!\gamma\!-\!\Lambda,3\Lambda\!-\!M_{\Lambda}\!-\!p\!+\!\gamma\!-\!2\alpha,\Lambda\!+\!\frac{M}{2}\!\bigg)\\
\times I\left(\lambda-\gamma,\gamma,\frac{\lambda+K}{2}\right)\frac{1}{\lambda+\mu-\gamma+L+1}\\
\times\sum_{\beta}(-1)^{\beta}\left(\!\!\begin{array}{c}
L-K \\
\beta \\
\end{array}\!\!\right)\left(\!\!\begin{array}{c}
L+K \\
L-M-\beta \\
\end{array}\!\!\right)\\
\times S\bigg(p+q-\gamma,L+\lambda-p+\mu-q,\\
\frac{\lambda-K}{2}+\mu+L-q-\Lambda-\frac{M}{2}-\beta\bigg),
\end{multline}
where $p$ and $q$ are related to $\epsilon$ and $\Lambda$ via Eqs.~(\ref{epsilon}) and~(\ref{Lambda}), and
\begin{align}
I(i,j,k)&=\sum_{n}(-1)^{n}\left(\!\!\begin{array}{c}
i \\
k-n \\
\end{array}\!\!\right)\left(\!\!\begin{array}{c}
j \\
n \\
\end{array}\!\!\right),\label{i}\\
S(i,j,k)&=\sum_{n}(-1)^{n}\left(\!\!\begin{array}{c}
i \\
n \\
\end{array}\!\!\right)\left(\!\!\begin{array}{c}
i+j \\
k+n \\
\end{array}\!\!\right)^{-1}.\label{s}
\end{align}
The inner product vanishes if $\Lambda+\frac{M}{2}$ is not integer. If the Elliott basis state is instead projected from the lowest-weight $\grpsu{3}\supset\grpu{1}\times\grpsu{2}$ basis state, the inner product is given by~(\ref{26}) with replacements $\lambda\to\mu$, $\mu\to\lambda$, $M_{\Lambda}\to-M_{\Lambda}$, $p\to\mu-q$, and $q\to\lambda-p$, which follows from state conjugation~\cite{draayer}.

The factors $I(i,j,k)$ and $S(i,j,k)$ appearing in~(\ref{26}) are precalculated using recurrence formulae~\cite{bahri}
\begin{equation}\label{i1}
I(i,j,k)=I(i-1,j-1,k)-I(i-1,j-1,k-2),
\end{equation}
for $i\ge j$, and
\begin{align}
S(i,j,k)\!&=\!S(i\!-\!1,j\!+\!1,k)\!-\!S(i\!-\!1,j\!+\!1,k\!+\!1),\label{s1}\\
S(i,j,k)&=\frac{(j-k)S(i,j-1,k)+iS(i-1,j,k)}{i+j},\label{s2}
\end{align}
with initial conditions
\begin{align}
I(i,0,k)&=\left(\!\!\begin{array}{c}
i \\
k \\
\end{array}\!\!\right),\label{i2}\\
S(0,j,k)&=\left(\!\!\begin{array}{c}
j \\
k \\
\end{array}\!\!\right)^{-1}.\label{i3}
\end{align}
For $i<j$, the relation~\cite{bahri}
\begin{equation}\label{isym}
I(i,j,k)=(-1)^{k}I(j,i,k)
\end{equation}
is used.

\subsection{Formulae for $\grpsu{3}\supset\grpu{1}\times\grpsu{2}$ reduced coupling coefficients}\label{appendix:canonical}

From the coefficients~(\ref{step1}), the coefficients~(\ref{step2}) are generated using iteratively the relation
\begin{multline}\label{21}
\rwig{(\lambda_{1},\mu_{1})}{\epsilon_{1}+3,\Lambda_{1}}{(\lambda_{2},\mu_{2})}{\epsilon_{2}\Lambda_{2}}{(\lambda_{3},\mu_{3})}{\epsilon_{3}^{\rm H},\Lambda_{3}^{\rm H}}_{\rho}\\
=\rbrajh{(\lambda_{1},\mu_{1})}{\epsilon_{1}+3,\Lambda_{1}}C^{(1,1)}_{+3,\frac{1}{2}}\rketjh{(\lambda_{1},\mu_{1})}{\epsilon_{1}\Lambda_{1}'}^{-1}\\
\times\sum_{\Lambda_{2}'=\Lambda_2\pm\frac{1}{2}}(-1)^{\Lambda_1-\Lambda_1'+\frac{1}{2}}\sqrt{\frac{2\Lambda_1+1}{2\Lambda_1'+1}}\\
\times U\left(\Lambda_3^{\rm H}\Lambda_2'\Lambda_1\frac{1}{2};\Lambda_1'\Lambda_2\right)\\
\times\rbrajh{(\lambda_{2},\mu_{2})}{-\epsilon_{2}\Lambda_{2}}C^{(1,1)}_{+3,\frac{1}{2}}\rketjh{(\lambda_{2},\mu_{2})}{-\epsilon_{2}-3,\Lambda_{2}'}\\
\times\rwig{(\lambda_{1},\mu_{1})}{\epsilon_{1}\Lambda_{1}'}{(\lambda_{2},\mu_{2})}{\epsilon_{2}+3,\Lambda_{2}'}{(\lambda_{3},\mu_{3})}{\epsilon_{3}^{\rm H}\Lambda_{3}^{\rm H}}_{\rho},
\end{multline}
where $\Lambda_1'=\Lambda_1\pm\frac{1}{2}$, and analytic expressions for the generator RMEs and $\grpsu{2}$ recoupling coefficients are available, \textit{e.g.}, in Refs.~\cite{hecht,hecht-vcs} and~\cite{var}, respectively. The relation~(\ref{21}) can be obtained from~(\ref{19}) by choosing $\epsilon_{1}\Lambda_{1}$ of the highest weight, which forces the first sum to vanish, and using the symmetry property
\begin{multline}\label{sympro}
\rwig{(\lambda_1,\mu_1)}{\epsilon_1,\Lambda_1}{(\lambda_2,\mu_2)}{\epsilon_2,\Lambda_2}{(\lambda_3,\mu_3)}{\epsilon_3,\Lambda_3}_{\rho}\\
=(-1)^{\Lambda_1-\Lambda_3+\varphi+\frac{\lambda_2-\mu_2}{3}-\frac{\epsilon_2}{6}}\sqrt{\frac{(2\Lambda_1+1)\dim(\lambda_3,\mu_3)}{(2\Lambda_3+1)\dim(\lambda_1,\mu_1)}}\\
\times\rwig{(\lambda_3,\mu_3)}{\epsilon_3,\Lambda_3}{(\mu_2,\lambda_2)}{-\epsilon_2,\Lambda_2}{(\lambda_1,\mu_1)}{\epsilon_1,\Lambda_1}_{\rho},
\end{multline}
where $\varphi=\lambda_{1}+\lambda_{2}-\lambda_{3}+\mu_{1}+\mu_{2}-\mu_{3}$, and
\begin{equation}\label{rdim}
\textrm{dim}(\lambda,\mu)=\frac{1}{2}(\lambda+1)(\mu+1)(\lambda+\mu+2)
\end{equation}
is the dimension of the irrep ($\lambda,\mu$) of $\grpsu{3}$.

From the coefficients~(\ref{step3}), the coefficients~(\ref{step4}) are generated using iteratively the relation\footnote{The relation~(\ref{18}) corresponds to~(18) of Ref.~\cite{draayer}, where, in the third of the four equations giving values for $X$, the expression $X\left(\Lambda_1-\frac{1}{2},\Lambda_2-\frac{1}{2}\right)$ on the left-hand side should be $X\left(\Lambda_1+\frac{1}{2},\Lambda_2-\frac{1}{2}\right)$.}
\begin{multline}\label{18}
\rwig{(\lambda_{1},\mu_{1})}{\epsilon_{1}\Lambda_{1}}{(\lambda_{2},\mu_{2})}{\epsilon_{2}+3,\Lambda_{2}}{(\lambda_{3},\mu_{3})}{\epsilon_{3}^{\rm H}\Lambda_{3}^{\rm H}}_{\rho}\\
=\rbrajh{(\lambda_{2},\mu_{2})}{\epsilon_{2}+3,\Lambda_{2}}C^{(1,1)}_{+3,\frac{1}{2}}\rketjh{(\lambda_{2},\mu_{2})}{\epsilon_{2}\Lambda_{2}'}^{-1}\\
\times\sum_{\Lambda_{1}'=\Lambda_1\pm\frac{1}{2}}(-1)^{\Lambda_1'-\Lambda_1+\frac{1}{2}}\sqrt{\frac{2\Lambda_2+1}{2\Lambda_2'+1}}\\
\times U\left(\Lambda_3^{\rm H}\Lambda_1'\Lambda_2\frac{1}{2};\Lambda_2'\Lambda_1\right)\\
\times\rbrajh{(\lambda_{1},\mu_{1})}{-\epsilon_{1}\Lambda_{1}}C^{(1,1)}_{+3,\frac{1}{2}}\rketjh{(\lambda_{1},\mu_{1})}{-\epsilon_{1}-3,\Lambda_{1}'}\\
\times\rwig{(\lambda_{1},\mu_{1})}{\epsilon_{1}+3,\Lambda_{1}'}{(\lambda_{2},\mu_{2})}{\epsilon_{2}\Lambda_{2}'}{(\lambda_{3},\mu_{3})}{\epsilon_{3}^{\rm H}\Lambda_{3}^{\rm H}}_{\rho},
\end{multline}
where $\Lambda_2'=\Lambda_2\pm\frac{1}{2}$, and analytic expressions for the generator RMEs and $\grpsu{2}$ recoupling coefficients are available, \textit{e.g.}, in Refs.~\cite{hecht,hecht-vcs} and~\cite{var}, respectively. The relation~(\ref{18}) can be obtained from~(\ref{19}) by choosing $\epsilon_{2}\Lambda_{2}$ of the highest weight, which forces the second sum to vanish, and using the symmetry properties~(\ref{sympro}) and
\begin{multline}
\rwig{(\lambda_1,\mu_1)}{\epsilon_1,\Lambda_1}{(\lambda_2,\mu_2)}{\epsilon_2,\Lambda_2}{(\lambda_3,\mu_3)}{\epsilon_3,\Lambda_3}_{\rho}\\
=\sum_{\rho'}\Phi_{\rho\rho'}[(\lambda_1,\mu_1),(\lambda_2,\mu_2);(\lambda_3,\mu_3)](-1)^{\Lambda_3-\Lambda_2-\Lambda_1}\\
\times\rwig{(\lambda_2,\mu_2)}{\epsilon_2,\Lambda_2}{(\lambda_1,\mu_1)}{\epsilon_1,\Lambda_1}{(\lambda_3,\mu_3)}{\epsilon_3,\Lambda_3}_{\rho'},
\end{multline}
where $\Phi_{\rho\rho'}[(\lambda_1,\mu_1),(\lambda_2,\mu_2);(\lambda_3,\mu_3)]$ is a “phase matrix" defined in terms of $Z$ recoupling coefficients (see Ref.~\cite{escher}).

\subsection{Formulae for $\grpsu{3}$ recoupling coefficients}

The system of linear equations used to calculate the $U$ coefficients is obtained from the system of equations~(\ref{22}) by fixing $\epsilon_{1}\Lambda_{1}$ and $\epsilon\Lambda$ to be of the highest weight and using the symmetry property~(\ref{sympro}). The resulting sytem of equations is
\begin{multline}\label{22m}
\sum_{\rho_{1,23}}\rwig{(\lambda_{1},\mu_{1})}{\epsilon_{1}^{\rm H}\Lambda_{1}^{\rm H}}{(\lambda_{23},\mu_{23})}{\epsilon_{23}\Lambda_{23}}{(\lambda,\mu)}{\epsilon^{\rm H}\Lambda^{\rm H}}_{\rho_{1,23}}\\
\times U[(\lambda_{1},\mu_{1})(\lambda_{2},\mu_{2})(\lambda,\mu)(\lambda_{3},\mu_{3});\\
(\lambda_{12},\mu_{12})\rho_{12},\rho_{12,3}(\lambda_{23},\mu_{23})\rho_{23},\rho_{1,23}]\\
=\sqrt{\frac{\textrm{dim}(\lambda_{12},\mu_{12})(\lambda_{1}+1)}{\textrm{dim}(\lambda_{1},\mu_{1})(2\Lambda_{12}+1)}}\\
\times\sum_{\substack{\epsilon_{2}(\epsilon_3\epsilon_{12})\\\Lambda_{2}\Lambda_{3}\Lambda_{12}}}\rwig{(\lambda_{12},\mu_{12})}{\epsilon_{12}\Lambda_{12}}{(\mu_{2},\lambda_{2})}{-\epsilon_{2}\Lambda_{2}}{(\lambda_{1},\mu_{1})}{\epsilon_{1}^{\rm H}\Lambda_{1}^{\rm H}}_{\rho_{12}}\\
\times\rwig{(\lambda_{12},\mu_{12})}{\epsilon_{12}\Lambda_{12}}{(\lambda_{3},\mu_{3})}{\epsilon_{3}\Lambda_{3}}{(\lambda,\mu)}{\epsilon^{\rm H}\Lambda^{\rm H}}_{\rho_{12,3}}\\
\times\rwig{(\lambda_{2},\mu_{2})}{\epsilon_{2}\Lambda_{2}}{(\lambda_{3},\mu_{3})}{\epsilon_{3}\Lambda_{3}}{(\lambda_{23},\mu_{23})}{\epsilon_{23}\Lambda_{23}}_{\rho_{23}}\\
\times(-1)^{\varphi+\frac{\lambda_{2}-\mu_{2}}{3}-\frac{\epsilon_{2}}{6}+\Lambda_{12}-\frac{\lambda_{1}}{2}}U\left(\frac{\lambda_{1}}{2}\Lambda_{2}\frac{\lambda}{2}\Lambda_{3};\Lambda_{12}\Lambda_{23}\right).
\end{multline}

The system of linear equations used to calculate the $Z$ coefficients is obtained from the system of equations~(\ref{mil}) by fixing $\epsilon_{13}\Lambda_{13}$ and $\epsilon\Lambda$ to be of the highest weight. The resulting sytem of equations is
\begin{multline}\label{z}
\sum_{\rho_{13,2}}\rwig{(\lambda_{13},\mu_{13})}{\epsilon_{13}^{\rm H}\Lambda_{13}^{\rm H}}{(\lambda_{2},\mu_{2})}{\epsilon_{2}\Lambda_{2}}{(\lambda,\mu)}{\epsilon^{\rm H}\Lambda^{\rm H}}_{\rho_{13,2}}\\
\times Z[(\lambda_{2},\mu_{2})(\lambda_{1},\mu_{1})(\lambda,\mu)(\lambda_{3},\mu_{3});\\
(\lambda_{12},\mu_{12})\rho_{12},\rho_{12,3}(\lambda_{13},\mu_{13})\rho_{13},\rho_{13,2}]\\
=\sum_{\substack{\epsilon_{1}(\epsilon_3\epsilon_{12})\\\Lambda_{1}\Lambda_{3}\Lambda_{12}}}\rwig{(\lambda_{1},\mu_{1})}{\epsilon_{1}\Lambda_{1}}{(\lambda_{3},\mu_{3})}{\epsilon_{3}\Lambda_{3}}{(\lambda_{13},\mu_{13})}{\epsilon_{13}^{\rm H}\Lambda_{13}^{\rm H}}_{\rho_{13}}\\
\times\rwig{(\lambda_{1},\mu_{1})}{\epsilon_{1}\Lambda_{1}}{(\lambda_{2},\mu_{2})}{\epsilon_{2}\Lambda_{2}}{(\lambda_{12},\mu_{12})}{\epsilon_{12}\Lambda_{12}}_{\rho_{12}}\\
\times\rwig{(\lambda_{12},\mu_{12})}{\epsilon_{12}\Lambda_{12}}{(\lambda_{3},\mu_{3})}{\epsilon_{3}\Lambda_{3}}{(\lambda,\mu)}{\epsilon^{\rm H}\Lambda^{\rm H}}_{\rho_{12,3}}\\
\times(-1)^{\Lambda_{1}+\frac{\lambda}{2}-\Lambda_{12}-\frac{\lambda_{13}}{2}}U\left(\Lambda_{2}\Lambda_{1}\frac{\lambda}{2}\Lambda_{3};\Lambda_{12}\frac{\lambda_{13}}{2}\right).
\end{multline}

\bibliographystyle{spphys}

\end{document}